\documentclass[aps,prx,twocolumn,nofootinbib,superscriptaddress,longbibliography,citeautoscript]{revtex4-1}

\pdfoutput=1
\usepackage[english]{babel}
\usepackage{letltxmacro}
\usepackage{latexsym}
\LetLtxMacro{\ORIGselectlanguage}{\selectlanguage}
\makeatletter
\DeclareRobustCommand{\selectlanguage}[1]{%
  \@ifundefined{alias@\string#1}
    {\ORIGselectlanguage{#1}}
    {\begingroup\edef\x{\endgroup
       \noexpand\ORIGselectlanguage{\@nameuse{alias@#1}}}\x}%
}
\newcommand{\definelanguagealias}[2]{%
  \@namedef{alias@#1}{#2}%
}
\makeatother
\definelanguagealias{en}{english}
\definelanguagealias{English}{english}
\usepackage{graphicx}
\usepackage{amsmath}
\usepackage{amsfonts}
\usepackage{amssymb}
\usepackage{bm}
\usepackage{color}
\usepackage[percent]{overpic}
\usepackage{soul} 
\usepackage{amssymb}
\usepackage{wasysym}
\usepackage{dsfont}
\usepackage{float}
\usepackage{hyperref}
\usepackage{verbatim}

\usepackage{hyperref}
\hypersetup{
    unicode=false,          
    pdftoolbar=false,        
    pdfmenubar=true,        
    pdffitwindow=false,     
    pdfstartview={FitH},    
    pdftitle={QMBS from periodic dynamics of entanglement degrees of freedom},    
    pdfauthor={W. W. Ho},     
    pdfsubject={},   
    pdfcreator={},   
    pdfproducer={}, 
    pdfkeywords={quantum scars} {ETH} {ergodicity}{thermalization}, 
    pdfnewwindow=true,      
    colorlinks=true,       
    linkcolor=black,          
    citecolor=blue,        
    filecolor=magenta,      
    urlcolor=blue           
}

\newcommand{\be}{\begin{equation}}
\newcommand{\ee}{\end{equation}}
\newcommand{\bea}{\begin{eqnarray}}
\newcommand{\eea}{\end{eqnarray}}

\renewcommand{\vec}[1]{\boldsymbol{\mathbf{#1}}}

\usepackage{graphicx}
\usepackage[colorinlistoftodos]{todonotes}
\usepackage{verbatim}
\usepackage[normalem]{ulem}

\newcommand{\braket}[2]{\mbox{$ \langle #1 | #2 \rangle $}}

\newcommand{\ket}[1]{\mbox{$ | #1 \rangle $}}
\newcommand{\bra}[1]{\mbox{$ \langle #1 | $}}

\newcommand{\Tr}{\mathrm{Tr}}

\newcommand{\iu}{{i\mkern1mu}}
\usepackage{amsmath,amssymb}
\newcommand{\updownarrows}{\uparrow\mathrel{\mspace{-1mu}}\downarrow}
\newcommand{\downuparrows}{\downarrow\mathrel{\mspace{-1mu}}\uparrow}
\renewcommand{\upuparrows}{\uparrow\uparrow}
\renewcommand{\downdownarrows}{\downarrow\downarrow}

\begin{document}

\title{ 
 Quantum many-body scars from virtual entangled pairs
}
\author{Sambuddha Chattopadhyay}
\affiliation{Department of Physics, Harvard University, Cambridge, MA 02138, USA }

\author{Hannes Pichler}
\affiliation{Department of Physics, Harvard University, Cambridge, MA 02138, USA }
\affiliation{ITAMP, Harvard-Smithsonian Center for Astrophysics, Cambridge, MA 02138, USA }
\affiliation{Division of Physics, Mathematics and Astronomy, California Institute of Technology, Pasadena, CA 91125, USA}

\author{Mikhail D.~Lukin}
\affiliation{Department of Physics, Harvard University, Cambridge, MA 02138, USA }

\author{Wen Wei Ho}
\thanks{Corresponding author: \href{mailto:wenweiho@fas.harvard.edu}{wenweiho@fas.harvard.edu} }
\affiliation{Department of Physics, Harvard University, Cambridge, MA 02138, USA }

\date{\today}
\begin{abstract}
{ 
We study weak ergodicity breaking in a one-dimensional, non-integrable spin-1 XY model.
We construct for it an exact, highly excited eigenstate, which despite its large energy density, can be represented analytically by a finite bond-dimension matrix product state (MPS) with area-law entanglement.
Upon a quench to a finite Zeeman field, the state undergoes periodic dynamics with perfect many-body revivals, in stark contrast to other generic initial states which instead rapidly thermalize.
This dynamics can be completely understood in terms of the evolution of entangled virtual spin-1/2 degrees of freedom,  
which in turn underpin the presence of an extensive tower of strong-eigenstate thermalization hypothesis (ETH)-violating many-body eigenstates.
The resulting  quantum many-body scars are therefore of novel origin. Our results provide important analytical insights into the nature and entanglement structure of quantum many-body scars.   
}

\end{abstract}

\maketitle
 
\section{Introduction}

Recent  experimental progress in the engineering and  control of well-isolated synthetic quantum systems, including ultracold atoms \cite{Bloch1, Greiner1, Greiner2,Langen1},  trapped ions \cite{HuseMonroe1}, Rydberg atom arrays \cite{Lukin1}, and spin qubits \cite{Lukin2}, has allowed for quantitative studies of fundamental physical phenomena such as thermalization and ergodicity in closed many-body systems. 
In such systems, the eigenstate thermalization hypothesis (ETH) \cite{ETH1, ETH2, ETH4} gives a generic prescription of thermalizing quantum dynamics. 
Known exceptions to the ETH   include strongly disordered, many-body localized (MBL) and integrable systems \cite{baskoMBL, husepal, spa1, husenandogan, husenandrev, spa2, RevModPhys.91.021001, integrability1, Integrability2}, wherein an extensive number of conservation laws break the ergodic hypothesis.
The nature of ergodicity-breaking in these cases pertains to that of  a {\it strong} kind---a finite fraction of    energy eigenstates (potentially all)  violate the ETH. 
Contrary to the volume-law entanglement   expected in the ETH, these states display sub-extensive amounts of entanglement.

Recently, it was noted that ergodicity  can instead be violated in a {\it weak} manner. 
Such a scenario was highlighted by quench experiments with arrays of interacting Rydberg atoms \cite{Lukin1}: Certain special  product states were observed to exhibit surprising, anomalously slow thermalizing dynamics marked by long-lived, periodic revivals, despite other simple initial states rapidly thermalizing as expected in a strongly-interacting system.
Subsequent theoretical studies \cite{qmbs5, PhysRevB.98.155134} have uncovered that underlying such dynamics are so-called `quantum many-body scars' (QMBS), an extensive set of atypical, low-entanglement, ETH-violating  energy eigenstates with finite energy density, which coexist with an otherwise ergodic spectrum---named in analogy to quantum scars in the single-particle quantum chaos literature  \cite{heller1984, PeriodicOrbits}.
QMBS have by now been studied in various contexts. They have been obtained by the ``embedding'' of special states via local projectors \cite{qmbs1, qmbs2, qmbs9}, exactly constructed in the AKLT model \cite{qmbs3,qmbs4}, uncovered in a spin-$1$ XY model \cite{jqi}, and even connected to gauge theories \cite{dalmonte} and quantum Hall physics \cite{sanjay}.

\begin{figure}[t]
    \includegraphics[width = 0.47 \textwidth]{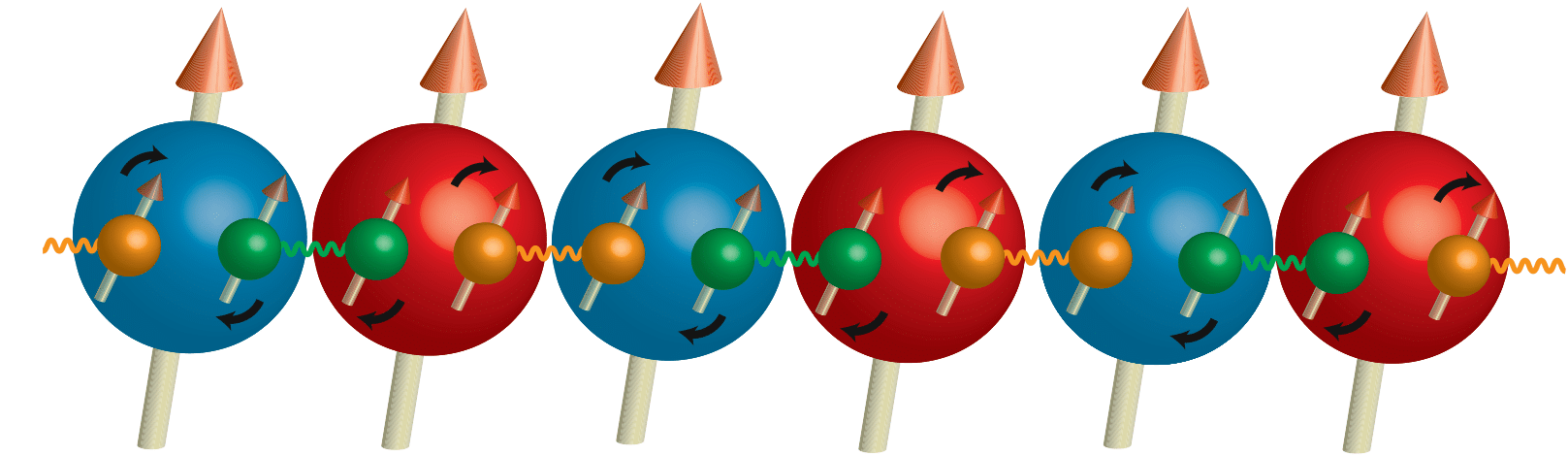}
\caption{ Quantum many-body scars from virtual entangled pairs. (a) Cartoon showing dynamics of matrix product state (MPS) of spin-1s (large red and blue spheres) in terms of rotation of underlying entangled virtual spin-1/2 pairs (small orange and green spheres). The periodic dynamics of the virtual entangled pairs yield dynamical recurrences of an entangled state at the physical level.
}
\label{fig:MPS_Cartoon}
\end{figure}

Despite intense efforts,  a general theory behind QMBS is still lacking and their origins have been vigorously debated \cite{PeriodicOrbits, qmbs8, qmbs9, qmbs10, qmbs11}. 
One signature for dynamical scarring is the presence of a su(2) algebra in the subspace of atypical eigenstates, leading to periodic dynamics of a ``large-spin" degree of freedom uncoupled to the rest of the system \cite{qmbs9}. Indeed, hidden, approximate  su(2) algebras have been found numerically in models hosting QMBS \cite{qmbs9,qmbs5}. 
Analytical models that display this phenomenon exactly are thus of great value for developing an understanding of the underlying mechanisms.
An important contribution in this context was provided by Ref.~\cite{qmbs9}, which gave a general recipe in constructing embedded su(2) algebras exactly in non-integrable toy models.
However, these examples are limited to simple representations of the algebra realized at the level of the physical constituent subsystems
--- thus, the ``large-spin'' is nothing more than a collective rotation of independent, unentangled local degrees of freedom. This means that the scarred trajectories obtained in this manner fundamentally do not contain any quantum entanglement, and their connection to scarred dynamics seen in experiments \cite{Lukin1}---which, in addition to periodic revivals, exhibit periodic entangling and disentangling of the atoms---is unclear. 
%
Developing an analytical understanding of QMBS beyond such simple models is thus crucial for a more comprehensive theory of this novel weak ergodicity-breaking phenomenon.   

In our work, we address this question by analytically finding QMBS in a spin chain where entanglement plays a crucial role. 
Concretely, we focus on a one-dimensional spin-$1$ XY model, construct a tower of QMBS as well as the corresponding  su(2) algebra, and argue that it is most easily understood in terms of underlying virtual, entangled degrees of freedom.
%
We first show that despite the non-integrable nature of the  model, we can write down a highly-excited, area-law entangled energy eigenstate, represented exactly as a bond-dimension $D=2$ matrix product state (MPS), which is made up of underlying virtual spin-1/2 entanglement degrees of freedom as caricatured in Fig.~\ref{fig:MPS_Cartoon}. Upon a quench to a finite Zeeman field,  this notably entangled state is driven out of equilibrium and undergoes perfectly periodic many-body revivals, in stark contrast to other highly out-of-equilibrium states which rapidly thermalize instead. 
The underlying, virtual spin-$1/2$ degrees of freedom provide direct insight into  this non-thermalizing dynamics: We find the state's unitary evolution can be  understood in terms of a collective rotation of pairs of entangled virtual spins, see Fig.~\ref{fig:MPS_Cartoon}. Underpinning these dynamics is therefore an $O(L)$ tower of provably lowly-entangled many-body eigenstates---quantum many body scars, where $L$ is the size of the system. In contrast to QMBS arising from the periodic dynamics of a   large spin belonging to an embedded su(2) algebra at the level of the physical degrees of freedom \cite{qmbs1,qmbs9},
%
%
%
our example  shows how the periodic dynamics of {\it virtual} entanglement degrees of freedom can also lead to QMBS but with fundamentally entangled scarred dynamics.

The mechanism developed in our work --- QMBS from ``virtual entangled pairs'' --- not only extends the known classes of analytically understood QMBS, it  also provides a way to understand entangled scarred dynamics (for example it could possibly offer insights into the entanglement oscillations observed in the Rydberg experiment \cite{Lukin1,qmbs5}), therefore providing a path towards a more complete theory of QMBS.
%
 


\section{Model}

To concretely illustrate our mechanism, we use our virtual entangled pair paradigm to explain entangled scarred dynamics in a spin-$1$ $XY$ quantum magnet model. We consider the following one-dimensional spin-1 XY Hamiltonian $H$ on $L$ sites with periodic boundary conditions: 
\begin{align}
\label{eq:H}
H & = H_\text{XY} + \epsilon V + h \sum_i S^z_i, \nonumber \\
H_ \text{XY}  &= J \sum_{i} \left( S^x_i S^x_{i+1} + S^y_i S^y_{i+1} \right),
\end{align}
where $S^\alpha$ ($\alpha = x,y,z$) are spin-$1$ operators acting on local states $|m\rangle \in \{ |1\rangle,|0\rangle,|-1\rangle \}$ satisfying $S^z|m\rangle = m |m\rangle$, and $\epsilon$ governs the strength of a perturbation $V = \sum_i (S_i^+)^2(S_{i+1}^-)^2 + \text{h.c.}$, where $S_i^{\pm} = S_i^{x} \pm i S_i^{y}$. The  Zeeman term $\sum_i S^z_i$  commutes with the Hamiltonian, and defines the magnetization sectors $M$. Additionally, the model has  spin-inversion, reflection, and translational symmetries. We hereafter set  $J=1$, $\epsilon = 0.2$, and work with even $L$.

\begin{figure}[t] 
    \includegraphics[width = 0.4 \textwidth, trim={0cm 0 0  0cm},clip]{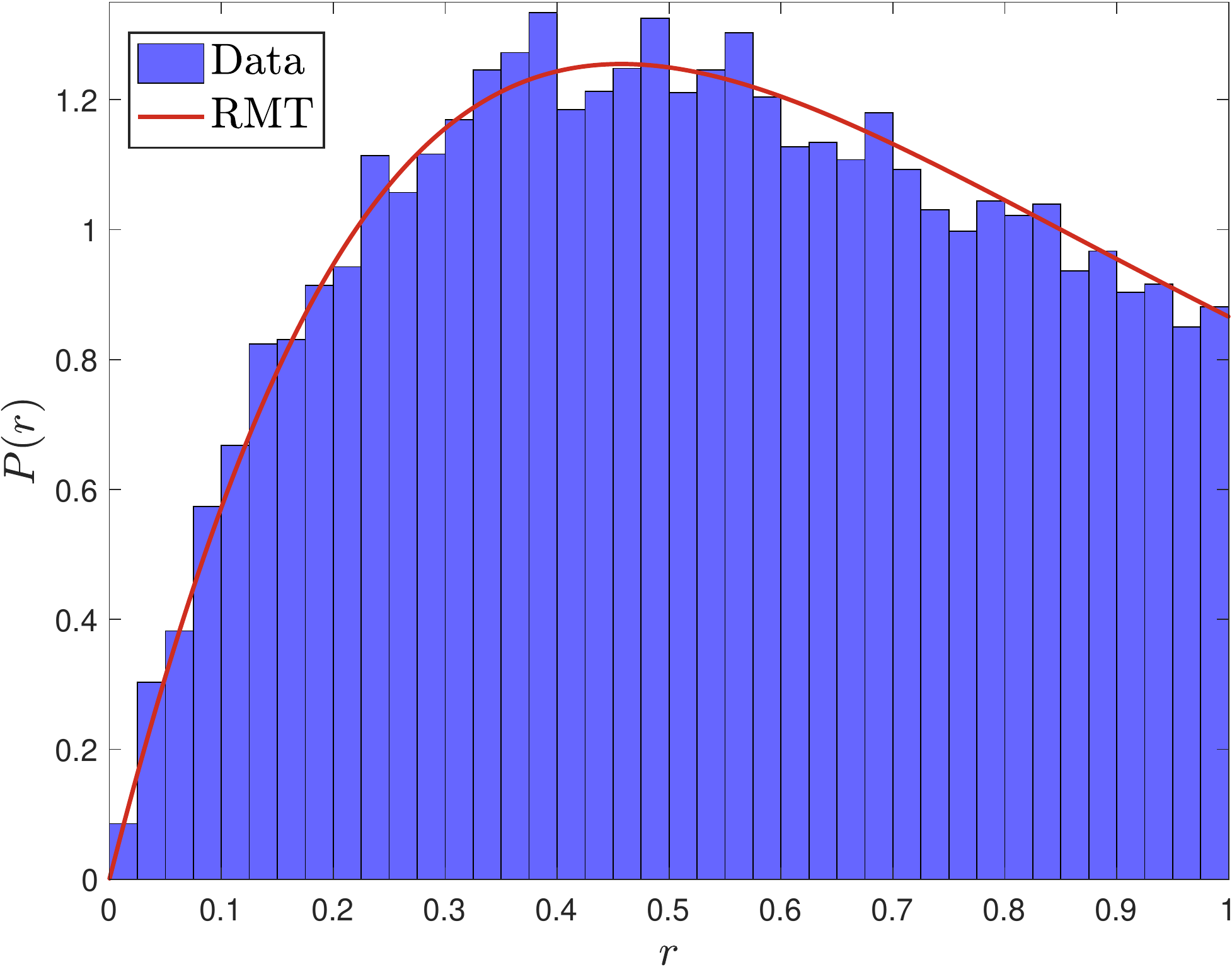}
\caption{Symmetry-resolved level spacing statistics of the Hamiltonian \eqref{eq:H} with $(J,\epsilon) = (1,0.2)$ (the value of $h$ is inconsequential). Distribution $P(r)$ in the momentum $k=0$, reflection $R=1$, and magnetization $M = -2$ sector  with dim($\mathcal{H}$)=18204, although similar results are obtained for any other $M$. The empirical distribution for $P(r)$ closely matches the analytic prediction (given for example in Eq. (9) of \cite{wwErg}) from the Gaussian Orthogonal Ensemble of random matrix theory (red), with average   $\langle r  \rangle = 0.53$.
}
  \label{fig:levelStatsPr}
\end{figure}

We add the perturbation $\epsilon V$ to render the model fully non-integrable: For $\epsilon = 0$ the system exhibits a peculiar behavior in which odd (even) magnetization sectors are ergodic (non-erogdic) due to a twisted SU(2) symmetry that exists only in the even sectors (for a detailed discussion see Appendix~\ref{App:A}) which is absent for $\epsilon \neq 0$. 
%
%
%
%
%
%
%
We probe this via the level spacing statistics:   upon resolving all possible  global symmetries as listed above and computing the distribution $P(r)$ of the level-spacing ratio $r_n = \frac{\min(\Delta E_n, \Delta E_{n+1})}{\max(\Delta E_n, \Delta E_{n+1})}$, where $\Delta E_n = E_{n+1} - E_n$ and $E_n$ is the ordered list of many-body energies \cite{husepal}), we find the level spacing statistics, in all symmetry-resolved sectors, approach the  prescriptions of the Gaussian Orthogonal Ensemble (GOE) of random matrix theory, indicating that the model is indeed non-integrable (see  Fig.~\ref{fig:levelStatsPr} for a representative example). 
%

Note that Ref.~\cite{jqi} considered a related spin-1 XY model and discovered a tower of  QMBS that can be exactly understood as arising from a particular  collective rotation of non-entangled physical spins, behaving as a `large-spin' of a  su(2) algebra representation realized simply at the physical level, 
in accordance with the framework of \cite{qmbs9}.
Additionally, Ref.~\cite{jqi} found numerical evidence for the existence of an additional tower of scars, termed ``bond-bimagnon'' scars, which origin was unknown as they fall outside the simple embedded algebra paradigm of \cite{qmbs9}. Our exact analytical analysis of the scars hosted by our model \eqref{eq:H} clarifies the nature of the latter tower of scars and in the process elucidates a novel paradigm for entangled scarred dynamics.

\section{An Exact Highly-Excited Eigenstate of the Spin-1 XY Model}

\subsection{MPS Representation}

Despite its non-integrable nature, the model \eqref{eq:H} harbors a special eigenstate $\ket{\psi_x}$ for $h=0$  which, while highly-excited (specifically it has zero-energy), is area-law entangled. This state can be represented as a bond-dimension $D=2$ periodic MPS with a two-site unit cell, 
\begin{align}
\ket{\psi_{x}} &= \sum_{m_1,   \dots, m_L} \text{Tr}(A_{m_1} B_{m_2} \cdots A_{m_{L-1}} B_{m_L})\ket{m_1, \cdots,m_L},
\label{eqn:MPS}
\end{align}
where $|m_i\rangle \in \{ |1\rangle_i,|0\rangle_i,|-1\rangle_i \}$ and the matrices $A_{m_i}$, $B_{m_i}$ are given by: 
\begin{align}
A_{\pm 1}=\frac{1}{\sqrt{2}}(1\mp \sigma^z)/2, A_0=\frac{1}{\sqrt{2}}\sigma^x, B_{m_i}=\sigma^z A_{m_i},
\end{align}
with $\sigma^x,\sigma^y,\sigma^z$ the standard Pauli matrices.
%
The MPS representation allows us to calculate the norm of the state \eqref{eqn:MPS} as $||\psi_x||=\sqrt{1+2(-\frac{1}{4})^{L/2}}$. Note that $\ket{\psi_x}$ is normalized in the thermodynamic limit (see Appendix~\ref{Appendix:MPSCalcs} for a comprehensive analysis).

A convenient way to verify that $\ket{\psi_x}$ is an eigenstate of \eqref{eq:H} is to show that the corresponding variance of the Hamiltonian, $\langle H^2 \rangle-\langle H \rangle^2$, vanishes identically. Using standard transfer matrix techniques for MPS we can indeed analytically compute the energy expectation value $\langle\psi_x| H|\psi_x \rangle$ and its second moment $\langle\psi_x| H^2|\psi_x \rangle$, both of which vanish, proving that $\ket{\psi_x}$ is a zero energy eigenstate. The technical steps of this calculation are lengthy but straightforward, and are presented in detail in Appendix~\ref{Appendix:MPSCalcs}.
%
By construction, the state $\ket{\psi_x}$ is short ranged entangled, obeying an area law. The von Neumann entanglement entropy (EE) $S_{vN}$ of a subsystem of $l$ contiguous sites approaches $\log(4)$ as $l \to \infty$ (see Appendix~\ref{Appendix:EE} for an explicit calculation). Note that this already implies a violation of the ETH so $|\psi_x\rangle$ is a QMBS of \eqref{eq:H} with $h=0$.
Furthermore, as the two-site transfer matrix has a single dominant eigenvalue, two-point correlation functions are  exponentially decaying. In particular, this implies that the MPS obeys the cluster decomposition $\lim_{|x-y|\to \infty} \langle O_x O_y \rangle - \langle O_x \rangle \langle O_y \rangle = 0$ for local $O_x, O_y$,  unlike the states such as Greenberger-Horne-Zeilinger  state which is a superposition of macroscopically different classical configurations.
%
We note also that the MPS is injective upon blocking of two sites that form a unit cell, implying that it can be prepared as the unique ground state of a local parent Hamiltonian \cite{Fannes:1992vq}. 

\begin{figure}[t]
    \includegraphics[width = 0.47 \textwidth]{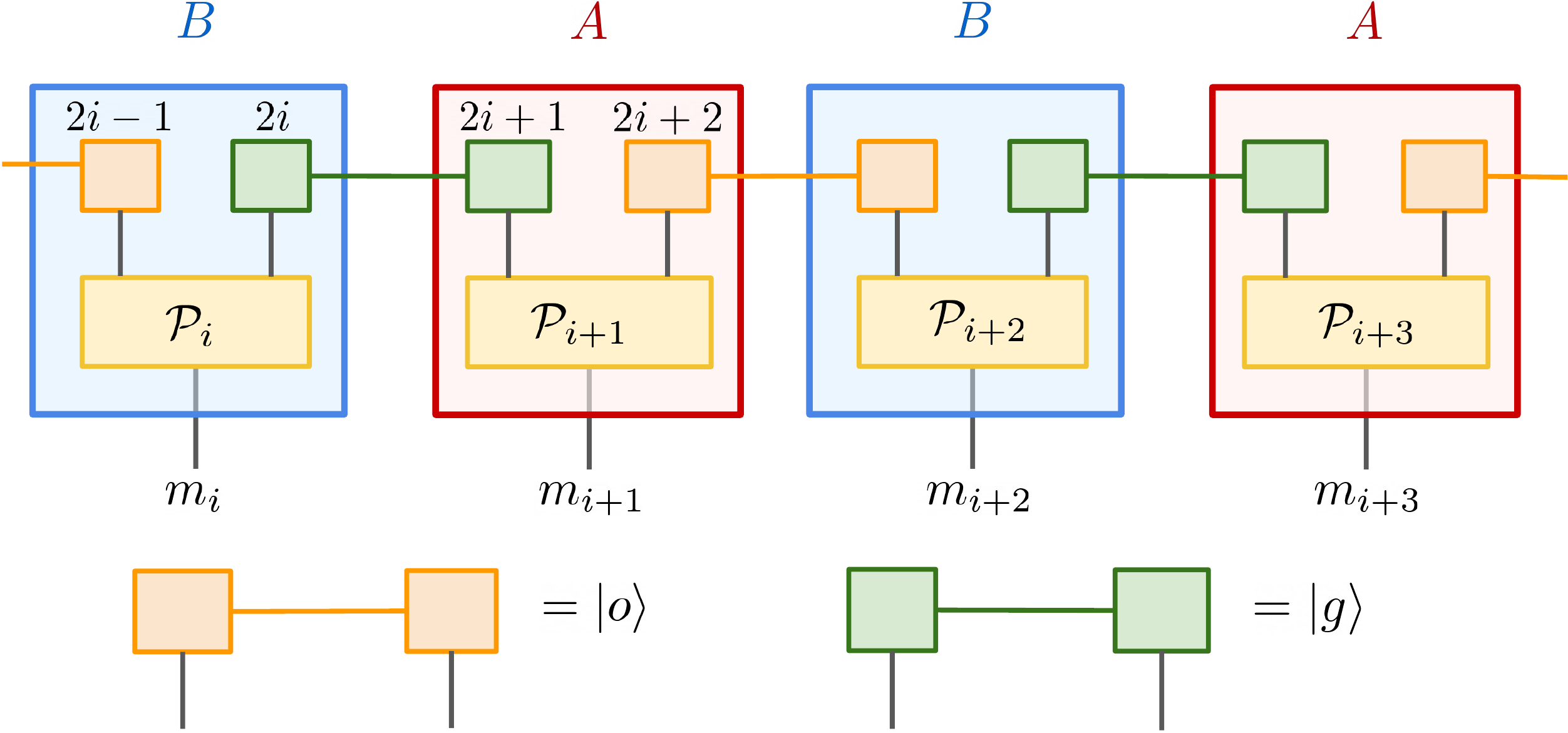}
    \caption{MPS representation of $\ket{\psi_x}$. The MPS  has $A$ and $B$ tensors,  constructed  by applying local maps $\mathcal{P}_i,$ given by Eq.~\eqref{eqn:localProj}, on pairs of spin-1/2s [e.g.~$(2i-1,2i)$, $(2i+1,2i+2)$]. The state of the spin-1/2s is given by an alternating pattern of locally entangled Bell pairs $\ket{o} = \frac{1}{\sqrt{2}} (\ket{\!\upuparrows}+\ket{\!\downdownarrows})$  and $\ket{g} = \frac{1}{\sqrt{2}} (\ket{\!\upuparrows}-\ket{\!\downdownarrows})$. The dangling legs of the MPS represent the state $m_i$ of the  $i$-th physical spin-1 system. 
    Note that the su(2) ladder operators \eqref{eqn:su2ops} act on pairs of virtual degrees of freedom e.g.~$(2i,2i+1)$ belonging to neighboring spin-1 systems.}
    \label{fig:MPS}
\end{figure}

%

\subsection{A virtual spin-$1/2$ construction}

As any MPS representation can be understood as a mapping from certain underlying virtual degrees of freedom to physical degrees of freedom, our MPS $\ket{\psi_x}$ can be constructed by projecting virtual spin-$1/2$ pairs onto physical spin-1 degrees of freedom. However in our case, elucidating such a ``spin-$1/2$ construction'' of the MPS  $\ket{\psi_x}$ will prove particularly valuable for  deciphering its intriguing non-thermalizing dynamics when the system is quenched to a finite field ($h \neq 0$) in the next section, Sec.~\ref{Sec:QMBS}. 


%

To begin the construction, consider replacing the physical spin-1 on site $i$  by two virtual spin-1/2s labeled $(2i-1,2i)$, so that there is in total a spin-1/2 chain of length $2L$, see Fig.~\ref{fig:MPS}.
Now entangle pairs of spin-1/2s belonging to two \textit{adjacent} spin-1s in an alternating fashion:
Place  the two spin-1/2s on sites $(4i,4i+1)$ (indices of the enlarged chain) in the Bell state  $\ket{o} = \frac{1}{\sqrt{2}} (\ket{\!\upuparrows}+\ket{\!\downdownarrows})$ 
and the two spin-1/2s on sites $(4i+2,4i+3)$ in an orthogonal Bell state  $\ket{g} = \frac{1}{\sqrt{2}} (\ket{\!\upuparrows}\!-\!\ket{\!\downdownarrows})$.
The MPS \eqref{eqn:MPS} is generated by mapping pairs $(2i-1,2i)$ of spin-1/2s of the virtual spin-1/2 chain to the spin-$1$ degrees of freedom at site $i$ on the physical chain via the local map
\begin{align}
 \mathcal{P}_i  & = \ket{1}_i\bra{\upuparrows\!\!}_{2i-1,2i} \! + \ket{0}_i(\bra{ \updownarrows}\!\!+\bra{\downuparrows\!\!})_{2i-1,2i} \! \nonumber \\
 & + \ket{\!-\!1}_i\bra{\downdownarrows\!\!}_{2i-1,2i},
\label{eqn:localProj}
\end{align}
  see Fig.~\ref{fig:MPS}. 

The $A$ and $B$ tensors defining the MPS representation can be recovered by contracting the open boundary MPS representations of $\ket{o}$ and $\ket{g}$ (in a suitable gauge) with the global map $\mathcal{P} = \bigotimes_i \mathcal{P}_i$ as shown in Fig.~3,
so that the MPS is obtained as
\begin{align}
|\psi_x\rangle = \mathcal{P} \left( \prod_i |o\rangle_{4i,4i+1} |g \rangle_{4i+2,4i+3} \right) \equiv \mathcal{P} |\phi_x \rangle,
\label{eqn:psix}
\end{align}
where $|\phi_x\rangle$ is the underlying spin-1/2 state  consisting of an alternating pattern of $|o\rangle$ and $|g\rangle$ Bell pairs.
Note that although this construction seems very similar to that of the AKLT ground state, the maps $\mathcal{P}_i$s we use are different.

\section{Quantum Many-body Scars from Virtual Entangled Pairs}

\subsection{Dynamical Signatures of Many-Body Scars}\label{Sec:QMBS}

%
\begin{figure}[tb]
  \includegraphics[width = 0.47 \textwidth]{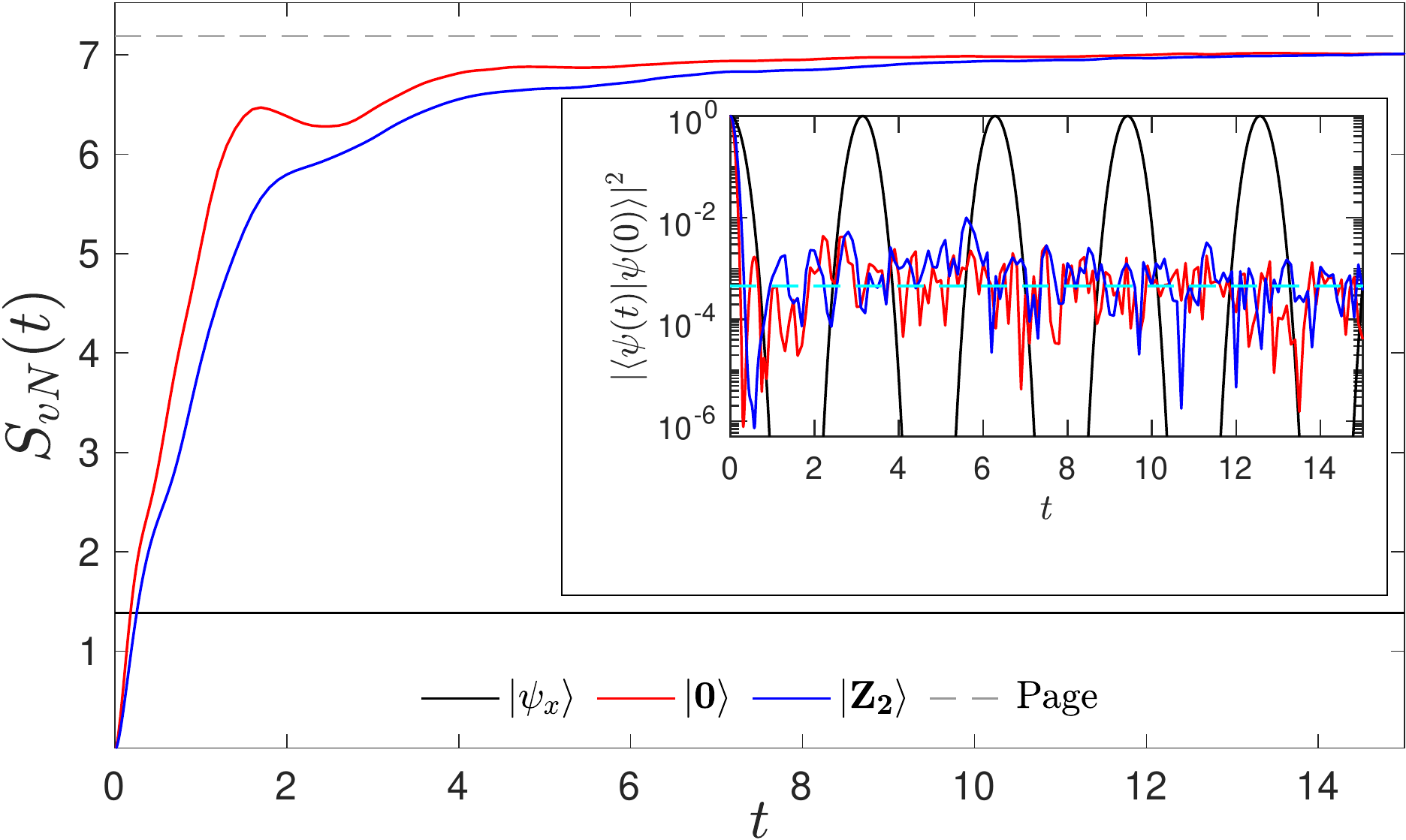}
\caption{ 
Quenches from various physical initial states with zero energies, obtained via exact diagonalization \cite{QuSpin1}  for   $(L,h, \epsilon) = (14,1, 0.2)$. (Main figure): Dynamics of half-chain bipartite von Neumann EE. $S_{vN}(t)$ of $\ket{\mathcal{\psi}_x}$ (black) remains finite and constant in time, while those of product states $\ket{\bf{\bm{Z}_2}}$ (blue) and $\ket{\textbf{0}}$ (red) grow rapidly and saturate near the Page value (dashed). 
(Inset): Return probabilities $|\langle \psi(t) |\psi(0)\rangle|^2$. 
$|\psi_x\rangle$ (black) revives periodically, perfectly, with return probability given asymptotically in the TDL by the expression $\cos^{2L}(ht)$, while that of $\ket{\bf{\bm{Z}_2}} $ (blue) and $\ket{\textbf{0}}$ (red)  quickly saturate and fluctuate around $1/\text{dim}(H)$ (cyan). 
}
  \label{fig:dynamics}
\end{figure}
While the existence of an exact area-law entangled, highly-excited eigenstate is itself interesting, our primary curiosity regarding $\ket{\psi_x}$ stems from its non-thermalizing dynamical behavior when quenched to a finite field ($h \neq 0$). Since $\ket{\psi_x}$ is a zero-energy eigenstate of $H$ with $h=0$, and $[H,\sum_i S_i^z]=0$, the unitary dynamics at finite $h$ is simply given by 
\begin{align}
    |\psi_x(t)\rangle = e^{-\iu H t}\ket{\psi_x} = \bigotimes_i \left[ e^{-\iu h S_i^{z} t} \right] \ket{\psi_x}.
    \label{eqn:dynamics}
\end{align}
That is, the dynamics simply reduces to a global rotation of all spins around the $z$-axis. This gives rise to perfectly periodic dynamics --- a common phenomenological theme of scarred dynamics \cite{heller1984, qmbs5,qmbs9, jqi}---with a period $T=2\pi/h$. Accordingly, local observables oscillate periodically and fail to approach thermal values at long times.  
Moreover, contrary to thermalizing expectations, $\ket{\psi_x}$'s entanglement entropy (for example measured by the von Neumann entanglement entropy of a subsystem) remains constant, as the unitary time evolution operator effectively factorizes into a product of local unitaries.
However, crucially, as $|\psi_x\rangle$ has non-zero entanglement to begin with, this implies that $|\psi_x(t)\rangle$ has also {\it finite}, non-zero entanglement in time --- a fundamentally entangled scar trajectory, see Fig.~\ref{fig:dynamics}.
%
%

%

Note that these observations are in stark contrast to the dynamics of generic weakly-entangled initial states, where the above reduction to a global rotation fails. As representative examples, we show numerical results for the dynamics two product states with similar energy densities as $\ket{\psi_x}$, that is $\ket{\bm{0}} = |0000\cdots\rangle$ and $\ket{\bm{Z}_2} = |-1,1,-1,1,\cdots\rangle$. Both those states do thermalize (Fig.~\ref{fig:dynamics}): Their half-chain entanglement entropies rapidly grow until they saturate near the Page value (of a random vector), and their return probabilities $|\langle \psi(t) |\psi(0)\rangle|^2$ decay quickly to the inverse of the Hilbert space dimension $\sim1/\dim (H)$.

\subsection{A virtual su(2) algebra underlying the tower of scars}
\label{sec:virtualsu2}

The presence of perfectly periodic dynamics implies that there is a set of eigenstates of $H$   with equally spaced energies that support the motion of the MPS $|\psi_x\rangle$. From the lemma of \cite{qmbs9}, we can say that at least one of them will be a non-ergodic, high-energy eigenstate, in the sense that it has a large $\sim 1/L$ overlap with a lowly-entangled state, i.e.~it is a quantum many-body scar. Indeed we will show in this section that the non-thermalizing dynamics of $|\psi_x\rangle$ is in fact supported by an extensive, $O(L)$ tower of QMBS. Our representation of $\ket{\psi_x}$ in terms of underlying virtual spin-$1/2$s can be leveraged to explicitly construct them as well as rigorously understand their entanglement structure.


Specifically, we will identify a set of operators $(J^z, J^+, J^-)$ acting on the \textit{virtual} spin-$1/2$ level obeying a su(2) algebra $[J^+, J^-] = 2 J^z, 
[J^z, J^{\pm}] = \pm J^{\pm}$, of which the underlying spin-1/2 configuration $|\phi_x\rangle$ in Eq.~\eqref{eqn:psix}  is the highest-weight state, i.e.~$|\phi_x\rangle$ has maximum eigenvalue under the total spin operator $\vec{J}^2 := \frac{1}{2}(J^+ J^- + J^- J^+) + (J^z)^2) $ and $J^x = \frac{1}{2}(J^+ + J^-)$. Upon decomposing $|\phi_x\rangle$ into simultaneous eigenstates $|\phi_n\rangle$ of the $\vec{J}^2, J^z$ operators, the QMBS are then obtained via the application of the global map $|\psi_n\rangle \equiv \mathcal{P} |\phi_n\rangle$.
%

Concretely, let us define the following operators   which act on the spin-$1/2$ chain of length $2L$: 
\begin{align}
\begin{tabular}{ccc}
\label{eqn:su2ops}
$J^z = \frac{1}{2}\sum_{i=1}^{2L} s^z_{i}$;
&$J^{\pm} =  \sum_{i=1}^{L} (-1)^i (s^{\pm}_{2i}  s^{\pm}_{2i+1} )$,
\end{tabular}
\end{align}
where $s^\alpha = \frac{1}{2} \sigma^\alpha$ ($\sigma^\alpha$ are the canonical Pauli matrices), $\alpha = x,y,z$, and $s^\pm = s^x \pm i s^y$.  
One can readily verify that they obey the su(2) commutation relations and hence form a particular representation of the algebra.
%
%
Crucially, note that the operators $J^{\pm}$ are not the standard spin-raising(lowering) su(2) operators corresponding to usual global spin rotations---they are instead sums of local terms that act simultaneously on two spin-1/2s straddling a physical spin-1 degree of freedom (see Fig.~\ref{fig:MPS} for more clarification). This ``straddling structure''gives rise to the non-trivial entanglement that the MPS $|\psi_x\rangle$ possesses. We remark that su(2) operators here are reminiscent of $\eta$-pairing operators \cite{YangEta} appearing in  the so-called pseudospin SU(2) symmetry of Hubbard models.

The non-standard generators \eqref{eqn:su2ops}  we have  identified can be used to organize the full Hilbert space of the virtual spin-1/2 chain of $2L$ sites into states that transform under  irreducible representations (irreps) of the su(2) algebra. Consider first two spin-1/2s, e.g.~$(2i,2i+1)$, and the restriction of the action of the operators $J^z, J^\pm$ on these spins, which still form a representation of the algebra.
We find the four states on these sites can therefore  be organized, according to this particular representation, into the irreps
\begin{align}
    \frac{1}{2} \oplus 0 \oplus 0,
\end{align}
where one $0$ irrep is spanned by the state $|\!\uparrow \downarrow \rangle$, the other $0$ irrep by $|\!\downarrow \uparrow\rangle$, and the $\frac{1}{2}$ irrep spanned by the two states $|\!\uparrow \uparrow\rangle, |\!\downarrow \downarrow\rangle$, which we can identify as a {\it pseudo-spin}. 
The virtual entangled pairs $|o\rangle = \frac{1}{\sqrt{2}} (|\!\uparrow\uparrow\rangle + |\!\downarrow\downarrow\rangle)$ and $|g\rangle = \frac{1}{\sqrt{2}} (|\!\uparrow\uparrow\rangle - |\!\downarrow\downarrow\rangle)$ are therefore states on the Bloch sphere of this pseudospin that point either in the $+x$ or $-x$ direction (depending on the sign in front of the local term of $J^+$); note they are entangled Bell pairs.  
The full Hilbert space is then organized as
\begin{align}
    \bigotimes_{i=1}^L \left( \frac{1}{2} \oplus 0 \oplus 0 \right) = \frac{L}{2} \oplus ...,
\end{align}
so that there is a unique total spin quantum number $J = L/2$.

From this discussion, one can  immediately see that $|\phi_x\rangle$, an alternating pattern of $|o\rangle$s and $|g\rangle$s, is the unique eigenstate of the operator $J^x = \frac{1}{2}(J^+ + J^-)$ with eigenvalue $J^x = L/2$ and therefore carries total spin quantum number $J=L/2$ (of the operator $\vec{J}^2$ which has eigenvalues $J(J+1)$). In other words, it is the highest weight state of the spin algebra, as claimed. 
%
%
%
%
We can therefore  further decompose $|\phi_x\rangle$  into a linear combination of the $L+1$ eigenstates  of the $J^z$ operator  $|\phi_n\rangle = |J^z=n-L/2\rangle$, $n = 0, \cdots, L$, also with total spin quantum number $J= L/2$, 
i.e.~$|\phi_x\rangle = \sum_{n=0}^L c_n' |\phi_n\rangle$  with real coefficients $c_n' = \sqrt{\frac{1}{2^L} {L \choose n}}$. Note that $|\phi_n\rangle$ are nothing but the ``Dicke-states'' of the virtual pseudo-spins. 
%
%
%

Now, consider the states on the spin-1 chain obtained by projecting the virtual pseudospin Dicke-states back to the physical spin-1 space, $|\psi_n\rangle \equiv \mathcal{P}|\phi_n\rangle$.
Since the map $\mathcal{P}$ preserves magnetization between the virtual and physical levels, $\mathcal{P} \sum_{i=1}^{2L} s^z_i = \sum_{i=1}^L S^z_i \mathcal{P}$, we can immediately conclude that $|\psi_n\rangle$---which can be straightforwardly shown to all be  normalized in the TDL (Appendix \ref{Appendix:MPSCalcs})---have well defined total magnetization $M = n-L/2$. 
It is easy to show  that $|\psi_n\rangle$ are in fact exact eigenstates of the Hamiltonian \eqref{eq:H} for generic $h$ with eigenvalues $E_n = h(2n-L)$: This simply follows from the fact that $|\psi_x \rangle$ is a zero-energy eigenstate of \eqref{eq:H} at $h=0$, and that all $|\psi_n\rangle$ are orthogonal as they have different $M$ quantum numbers. 

Thus,   what we have shown is that the projected Dicke states $|\psi_n\rangle$ span an $O(L)$ degenerate zero-energy eigenspace of the Hamiltonian at $h=0$, which, when $h \neq 0$, splits into the tower of eigenstates of \eqref{eq:H} with equally spaced energies that underlies the non-thermalizing, periodic dynamics of $|\psi_x\rangle$.

%
%

%
We note that $|\psi_n\rangle$ obtained this way, is in one-to-one correspondence to the set of so-called ``bond-bimagnon'' states $|\mathcal{S}'_n\rangle$, given in the appendix of \cite{jqi}, conjectured and supported by numerical evidence to be scars of a closely related spin-1 XY Hamiltonian. Up to normalization, they are defined by
$
\ket{S_n^{'}}  \propto \sum_{i_1 \ne i_2 \ne i_3 \cdots \ne i_n} (-1)^{\sum_{k=1}^{n} i_k} \Pi_{k=1}^{n}  S^+_{i_k}S^+_{i_k+1} \ket{-1,-1,\cdots}
$
%
(recall that $S^+$ are spin-1 raising operators). Our present analysis clarifies the algebraic structure behind this set of states, and the next section elucidates their entanglement structure, rigorously establishing them as ETH-violating quantum scars.

Let us also note that the su(2) structure uncovered  yields a more microscopic understanding of the precise  trajectory of $|\psi_x(t)\rangle$ through the many-body Hilbert space. From $\mathcal{P} \sum_{i=1}^{2L} s^z_i = \sum_{i=1}^L S^z_i \mathcal{P}$, we get  from \eqref{eqn:dynamics} that $|\psi_x(t)\rangle = \mathcal{P} e^{-i h \sum_{i}^{2L} s^z_i t} |\phi_x\rangle$, where $|\phi_x\rangle$ is the underlying alternating pattern of $|o\rangle$s and $|g\rangle$s. Thus, we can interpret the dynamics as manifesting completely at the level of the virtual spin-1/2 level.
From this point of view, local virtual entangled pairs $\ket{o}$ or $\ket{g}$, which point in the $+x$ or $-x$ axis of their respective pseudospin Bloch spheres, rotate around the $z$-axis on the equatorial plane, tracing out a contour of maximal entanglement. 
Alternatively, when viewed globally, the periodic dynamics can be thought of as a precession of a large collection spin $|\phi_x\rangle$ of the su(2) algebra \eqref{eqn:su2ops} initially pointing in the $x$-direction around the $z$-axis, but  at the virtual spin-1/2 level.
%
%
%
Importantly, remembering to incorporate back the action of the map $\mathcal{P}$ to return to the physical spin space, the map induces nontrivial entanglement between the physical spin-1 degrees of freedom, preventing a simple decoupled description of rotation of physical spin-1s, unlike the general construction prescribed by \cite{qmbs9}. From this discussion, it can also be seen that the functional form of $|\psi_x(t)\rangle$'s  return probability $|\braket{\psi_x(t)}{\psi_x (0)}|^2 $ goes asymptotically in the TDL as $ \cos^{2L}{(h t)}$, as seen in Fig.~\ref{fig:dynamics}.

\subsection{Entanglement Structure of QMBS}
\begin{figure}[t]
    \includegraphics[width = 0.44 \textwidth]{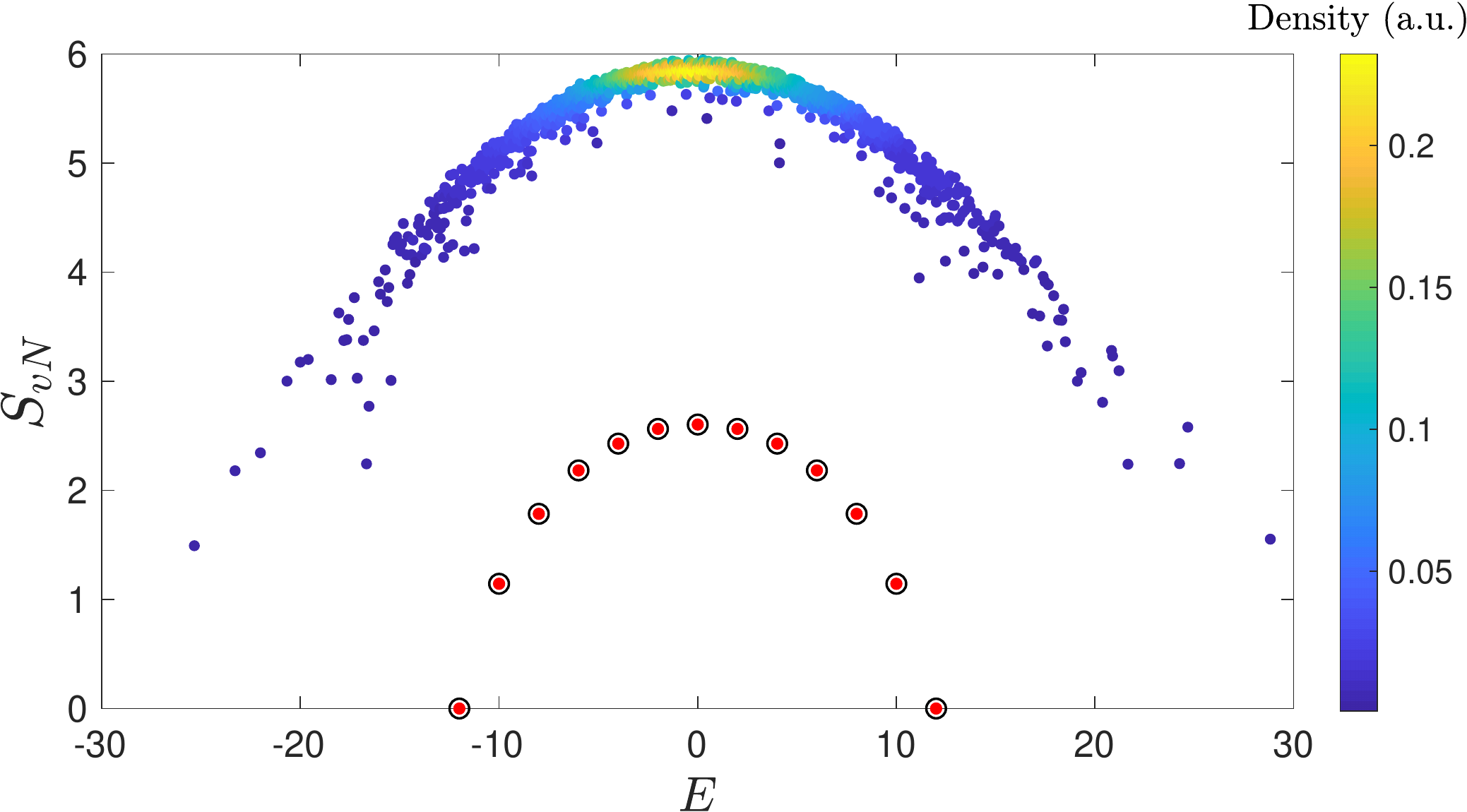}
\caption{Half-chain bipartite von Neumann EE of eigenstates of \eqref{eq:H} for $L=12$ and $(h, \epsilon) = (1,0.2)$. Smaller dots indicate zero-momentum, spin-inversion and reflection-symmetric states with  $M=0$, with   color referring to the local density of states. They form a highly-entangled branch, with expected volume-law scaling of EE from the ETH. A lowly-entangled branch of eigenstates, well separated from the bulk, appears below: These are the QMBS states (circled larger red dots, plotted for different $M$), with at most logarithmic scaling of EE.}
  \label{fig:SvE}
\end{figure}

The   su(2) structure on the entanglement degrees of freedom underlying $|\psi_n\rangle$, also allows us to rigorously establish that the eigenstates $|\psi_n\rangle$ are sub-thermally entangled eigenstates obeying an at most logarithmic  entanglement scaling law, and hence ETH-violating. In other words, $|\psi_n\rangle$ are, like $|\psi_x\rangle$, many-body scars. 

To see this, we first notice that as  $|\phi_n\rangle$ are Dicke states of pseudospins at the virtual spin-$1/2$ level, they have a well known scaling of EE. Specifically, consider a  bipartition of the spin-1/2 system into a contiguous region with $2l$ spin-1/2s and a region containing the rest of the $2L-2l$ spin-1/2s, where entanglement cuts, at each boundary, are made  between two spin-1/2s which comprise a single physical spin-1. Note that such a bipartition respects the local action of $J^\pm$ (i.e. a single $s_{2i} s_{2i+1}$ in $J^{\pm}$, see Eq.~\ref{eqn:su2ops}) which acts across neighboring physical spin-$1$s (see Sec.~\ref{sec:virtualsu2}). Thus, the contiguous region of $2l$ spin-1/2s encloses exactly $l$ pseudospins. Therefore, we can say that the von Neumann entanglement entropy of $|\phi_n\rangle$ obeys $S_{vN} = O(\log(l))$. 

This entanglement bipartition is, of course, not physically meaningful when viewed at the level of the spin-1 chain---to form such a bipartition, we would be attempting to `cut' physical spins-1s. 
%
If we instead consider appending the two spin-1/2s immediately neighboring the previous contiguous region, thereby forming a new region that encloses $2l+2$ spin-1/2s, then  the EE of this subregion can only change by at most $2\log(2)$, a fact which follows from the subadditivity of EE and the Araki-Lieb triangular inequality. 
%
   
%
%
This new entanglement bipartition  is now well-defined on the physical spin chain level, as one of the regions encompasses $l+1$ physical spin-1s and the other $L-l-1$ spin-1s. This allows us to directly compare the von Neumann entanglement entropies of the pseudospin states $|\phi_n\rangle$, as we have discussed above, with the spin-1 states $|\psi_n\rangle$ of the same subregion, as the two differ merely by the application of a product of local maps $\mathcal{P} = \bigotimes_i \mathcal{P}_i$ which respects the locality of this entanglement bipartition.
%
Specifically, we can invoke Nielsen's theorem \cite{Nielsen}, a result from quantum information theory which provides a general condition for when a pure state of a bipartite quantum system may be transformed into another using only so-called ``local operations and classical communication'' (LOCC) operations: namely, that it is  possible to do so if and only if the singular values in the Schmidt decomposition of the final state majorizes that of the initial state  \cite{footnote}. In particular, this implies that entanglement measures cannot increase via the LOCC operations.  
In the present case, as $|\psi_n\rangle$ is precisely a state obtained from $|\phi_n\rangle$ via the application of $\mathcal{P}$, an LOCC operation, its entanglement cannot be larger than that of $|\phi_n\rangle$'s and thus
%
we can conclude that $|\psi_n\rangle$ has von Neumann EE obeying $S_{vN} = O(\log(l))$.

%
%
In Fig.~$\ref{fig:SvE}$, we plot the half-chain bipartite von Neumann EE of the energy eigenstates of a chain of $L=12$ spin-1s in a representative symmetry-resolved sector, as well as  of the QMBS $\ket{\psi_n}$. 
We expect that as the system is non-integrable, the majority of states should obey an extensive (volume-law) scaling of von Neumann EE, according to ETH predictions. In particular, states near the middle of the spectrum should saturate the Page limit  $(L/2)\log(3) - 1/2$. This is indeed what we find -- bulk excited states form a highly entangled branch. In contrast, the states $|\psi_n\rangle$ are much less entangled, forming instead a lowly-entangled branch with equally spaced energies, well separated from the rest of the system. This corroborates the previous analytical considerations, firmly establishing the states $|\psi_n\rangle$ as ETH-violating eigenstates.

\section{Discussion \& Outlook}
While the phenomenology described above---an extensive set of lowly-entangled many-body eigenstates embedded in a chaotic spectrum (i.e.~ergodicity breaking, ETH-violating QMBS) as well as non-thermalizing dynamics   from certain special initial conditions manifested in  perfect periodic recurrences --- has been studied in previous models, the present system provides a first example of an analytically understood model where the oscillatory dynamics cannot simply be reduced to a decoupled description and is instead  fundamentally entangled. 
We showed that this is related to the precession of a `large' spin at the level of underlying entanglement degrees of freedom, evolving under a non-standard su(2) algebra which acts on \textit{pairs} of virtual sites; the underlying precession is then projected back to the physical spin level via a nontrivial map \footnote{In fact, one can show that the spin-1/2 raising/lowering operators $J^\pm$ we defined cannot be `pushed' to the physical level as spin-1 raising/lowering operators, at least at the level of local two-body operators.}.
%
%
%
This is in contrast to the scenarios where weak ergodicity breaking  has been understood analytically in terms of a collection of independently rotating---hence un-entangled---physical spins of a large global angular momentum sector, which are shielded from thermalization processes \cite{qmbs9, jqi}. In this respect, dynamics in the present model are closer to the periodic entangling and disentangling  dynamics of the PXP model of the Rydberg experiments \cite{Lukin1,qmbs5}.
%

Our work uncovers a novel mechanism involving the dynamical evolution of underlying entanglement degrees of freedom which give rise to scars, thereby adding to our general understanding of this weak ergodicity-breaking phenomenon.
It would be interesting to explore if there are other models that exhibit scars that lie within such a similar theoretical framework.
%
An immediate consideration is   higher-spin   models, for which an analogous construction of an AKLT-like  MPS and corresponding nontrivial algebra on the underlying entanglement degrees of freedom can be carried out. 
Our results may also yield some insights into the nature of the scarred trajectories  in the Rydberg simulator experiments \cite{Lukin1}, which do exhibit entanglement, as well as have connections to QMBS found in the AKLT-model \cite{qmbs3, qmbs4}. Finally,  connections to scars in lattice gauge theories \cite{dalmonte} can be explored using the present approach.

 \begin{acknowledgements} {\it Acknowledgments.} ---  
 We thank S.~Choi for useful discussions and M.~Bukov for help with the package QuSpin with which exact diagonalization studies were carried out.
 This work was supported through the National Science Foundation (NSF), the Center for Ultracold Atoms, DOE.~Office of Naval Research and the Vannevar Bush Fellowship. 
 S.C.~is supported by the Herschel Smith Undergraduate Program and the Jacob Wendell Scholarship Prize. H.P.~was supported by the NSF through a grant for the Institute for Theoretical Atomic, Molecular, and Optical Physics at Harvard University and the Smithsonian Astrophysical Observatory, and
 by the Gordon and Betty Moore Foundation’s EPiQS Initiative, Grant GBMF8682. W.W.H.~is supported by the  Moore Foundation's EPiQS Initiative   Grant No. GBMF4306 and the NUS Development Grant AY2019/2020. \end{acknowledgements}

\begin{appendix} 


 
 \section{Level statistics dependence in pure spin-1 $XY$-model: `Twisted' SU(2) symmetry in even magnetization sectors}
 \label{App:A}

We explain here the peculiar level spacing statistics dependence on the magnetization quantum number of the pure (i.e.~unperturbed) spin-1 $XY$-model in one-dimensions and periodic boundary conditions, given by:
\begin{align}
H_\text{XY} = J \sum_{i=1}^L   \left( S^x_i S^x_{i+1} + S^y_i S^y_{i+1} \right),
\end{align}
where $S^\alpha, \alpha = x,y,z,$ are spin-1 operators. $H_\text{XY}$ has translational symmetry, magnetization (given by $M = \sum_i S^z_i$) spin-inversion and reflection symmetry.
This model has magnetization, translation, spin-inversion, and reflection symmetries. 

%

%

\begin{figure}[t]
    \includegraphics[width = 0.4 \textwidth, trim={0cm 0 0  0cm},clip]{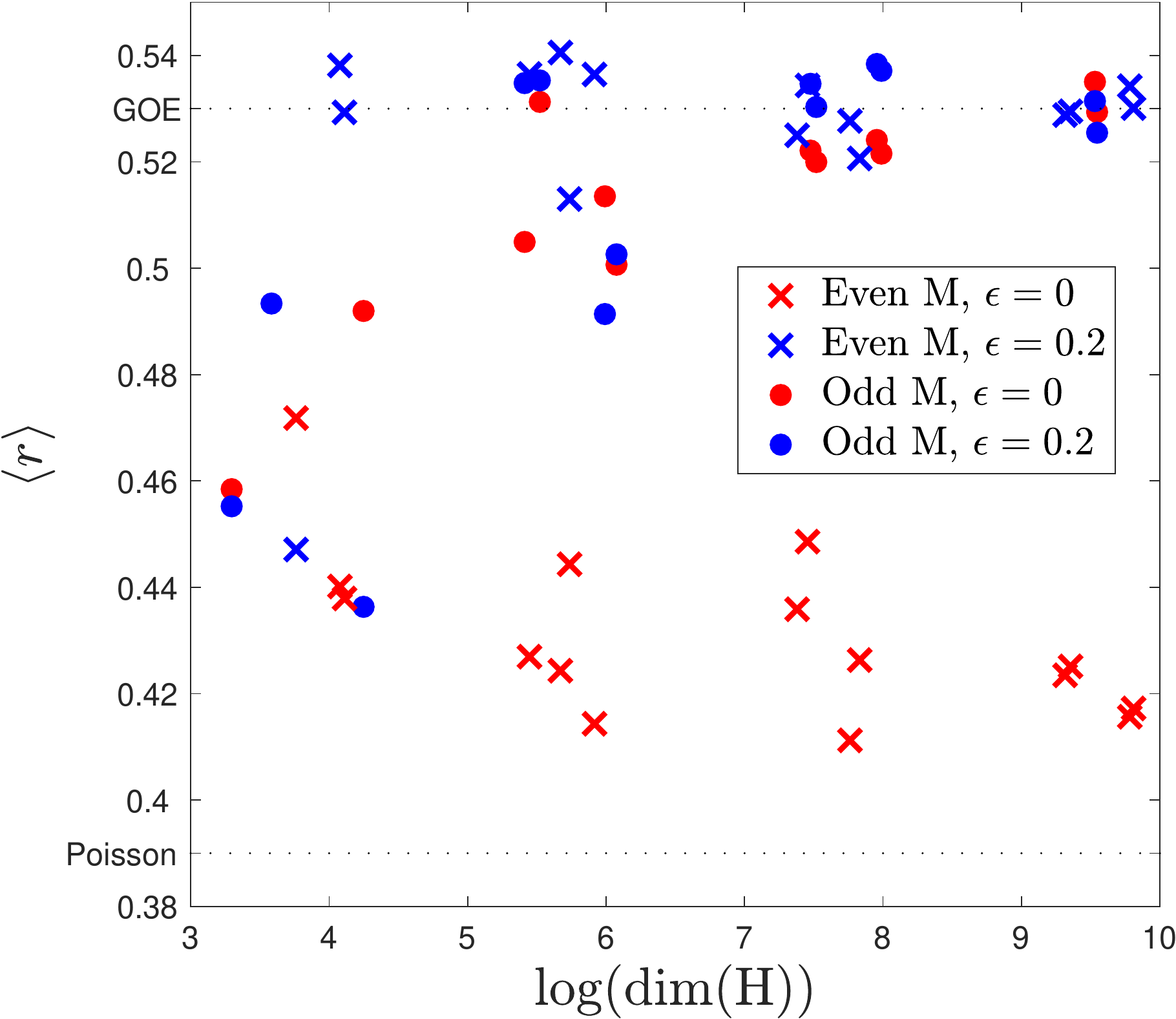}
\caption{Level statistics are contingent upon magnetization sector $M$. Symmetry-resolved $\langle r \rangle$. Data is for various $L$ and $M$ up to $L =14$ and momentum sectors $k=0,\pi$.
Red markers: Perturbation strength $\epsilon=0$. Odd $M$(circles) tend towards the Wigner-Dyson class belonging to the Gaussian Orthogonal Ensemble (GOE), with $\langle r \rangle \approx 0.53$. Even $M$(crosses) in contrast are inconsistent with GOE statistics.
Blue markers: Perturbation strength $\epsilon=0.2$. Odd[even] $M$ (circles[crosses]) now all tend towards GOE statistics.
%
}
  \label{fig:levelStats}
\end{figure}

Upon diagonalizing the Hamiltonian in different momentum and magnetization sectors, as well as resolving fully the remaining global symmetries, one finds for example that the $r$-level spacing statistics, defined by
\begin{align}
r_n = \frac{\min(\Delta E_n, \Delta E_{n+1} )}{\max(\Delta E_n, \Delta E_{n+1} )} \in [0, 1],
\end{align}
where $\Delta E_n = E_{n+1} - E_n$ for the sorted list of energies, tends, for large system sizes, to
\begin{align}
\langle r \rangle  \to
\begin{cases}
\sim 0.53 & \text{ for } M \text{ odd }, \nonumber \\
\neq 0.53 & \text{ for } M \text{ even },
\end{cases}
\end{align}
and $\langle \cdot \rangle$ denotes averaging.
For the former case, $\langle r \rangle \approx 0.53$ is consistent with that of Wigner-Dyson (WD) statistics in the Gaussian Orthogonal Ensemble (GOE), indicating the model in those sectors is non-integrable and chaotic.
 For the latter case, $\langle r \rangle$ tends neither towards a value expected of a WD-class, nor towards Poissonian statistics where $\langle r \rangle \approx 0.39$, but instead hovers around $\langle r \rangle \approx 0.4$, even in the thermodynamic limit (TDL), see Fig.~\ref{fig:levelStats}.

This indicates that in   even magnetization sectors, there are further unresolved symmetries.
Indeed, we are able to explain this peculiar behavior as arising from a {\it twisted} SU(2) symmetry, but which only affects the even magnetization sectors, so that the full Hamiltonian $H_\text{XY}$ does not   have the SU(2) symmetry.

Kitazawa et. al. studied the XY model in open boundary conditions and an ``artificial'' one, showing that the models possessed a twisted SU(2) symmetry \cite{su2symmetrytwisted}. Specifically, for the latter case, the Hamiltonian was 
\begin{align}
H_\text{XY}' & = \sum_{i=1}^{L-1} J_{i,i+1} \left( S^x_i S^x_{i+1} + S^y_i S^y_{i+1} \right) \nonumber  \\
& + \frac{J_{L,1}}{2} \left( S_L^+ S_1^- e^{\mp i \frac{\pi}{2} M } + S_L^- S_1^+ e^{\pm i \frac{\pi}{2} M }\right),
\end{align}
where $S^\pm = S^x \pm i S^y$ and $J_{ij}$ are arbitrary real coefficients.
Defining the operators 
\begin{align}
\tilde{s}^{\pm}_i = \frac{1}{2} \left( S^\pm_{i} \right)^2, \qquad \tilde{s}^z_i = \frac{1}{2} S^z_i,
\end{align}
one can show that they form a basis of an su(2) algebra (which is not the standard one)
\begin{align}
[\tilde{s}^z_i, \tilde{s}^\pm_j] = \pm \delta_{ij} \tilde{s}^\pm_i, \qquad [\tilde{s}^+_i, \tilde{s}^-_j] = 2 \delta_{ij} \tilde{s}^z_i.
\end{align}
Furthermore, one can also define the operators
\begin{align}
s^\pm_i = \tilde{s}^\pm_i U_i, \qquad s^z_i = \tilde{s}^z_i,
\end{align}
where 
\begin{align}
&  U_i = \prod_{l=1}^{i-1} \left( 1 - 2(S^z_l)^2 \right) = e^{i \pi \sum_{l=1}^{i-1}} s^z_i \text{ for } i > 1,
\end{align}
with $U_1 =\mathbb{I}$ and which, obey identical commutation relations as $\tilde{s}^\alpha_i$ with $\tilde{s}^\alpha_i \mapsto s^\alpha_i$. From these definitions it is possible to show, that the global operators
\begin{align}
 s^\pm_T &= \sum_{i=1}^L s^\pm_i,  \qquad s^z_T  = \sum_{i=1}^L s^z_i = \frac{1}{2} M 
\end{align}
also obey the commutation relation of su(2), i.e.~
\begin{align}
[ s^z_T, s^\pm_T] = \pm s^\pm_T, \qquad [s^+_T, s^-_T ] = 2s^z_T.
\end{align}

Now, through a lengthy but straightforward calculation as shown in \cite{su2symmetrytwisted} which we do not reproduce, one can derive that $H_\text{XY}'$ commutes with $s^z_T, s^\pm_T$. 
This implies that $H_\text{XY}'$ has an SU(2) symmetry, though not generated by the canonical spin-raising and lowering operators, but rather by $s^\alpha_T$, and so its energies and eigenstates are organized in representations of this ``twisted'' SU(2) algebra.
In other words, this model, while interacting, is integrable. We therefore do not expect its level statistics, even upon resolving all possible global symmetries (magnetization, translation, if it exists, etc.), to tend towards a WD class.

This result allows us to make a connection to the model we study, $H_\text{XY}$.
Note that $H_\text{XY}$ does not commute with $s^\pm_T, s^z_T$ and so does not possess the twisted SU(2) symmetry.
Nevertheless, its spectra in   even magnetization sectors {\it coincides} with that of {\it some} $H_\text{XY}'$; thus, its level spacing statistics in those sectors will {\it not} be of the WD class.
To see this, consider $M = 4n$ where $n$ integer.
Take $J_{i,i+1} = J$ for $ i = 1, \cdots, L-1$ and $J_{L,1} = J$ in $H_\text{XY}'$. Then, we see that
\begin{align}
\text{spectrum}(H_\text{XY}) = \text{spectrum}(H_\text{XY}') \text{ for } M = 4n.
\end{align}
On the other hand, consider $M = 4n+2$ where $n$ integer. 
Take $J_{i,i+1} = J$ for $ i = 1, \cdots, L-1$ and $J_{L,1} = -J$ in $H_\text{XY}'$. Then, we see that
\begin{align}
\text{spectrum}(H_\text{XY}) = \text{spectrum}(H_\text{XY}') \text{ for } M = 4n+2.
\end{align}
This explains the observed trend in level spacing statistics of $H_\text{XY}$.

The presence of $V = \sum_i (S^+_i)^2(S^-_{i+1})^2 + (S^-_i)^2 (S^+_{i+1})^2$ (as used in the main text) removes such dependencies and makes the level statistics of all symmetry-resolved sectors obey   Wigner-Dyson statistics, see Fig.~\ref{fig:levelStats} when the perturbation strength $\epsilon \neq 0$, whilst preserving the condition the MPS $|\psi_x\rangle$ is a zero-energy eigenstate. Besides this term, we also find that the terms
\begin{align*}
V' &=i\sum_{i}(S^+_i)^2(S^-_{i+1})^2 - (S^-_i)^2 (S^+_{i+1})^2, \\ 
V''&=\sum_{i} (S^z_i)^2 S^z_{i+1} - S^z_i (S^z_{i+1})^2, \\
V''' &= \sum_{i} (S^z_i)^2 (S^z_{i+1})^2 - S^z_i S^z_{i+1}
\end{align*}
 have similar effect.

\section{Calculations using the Matrix Product State  $\ket{\psi_x}$}
\label{Appendix:MPSCalcs}

In this section, we use the MPS representation of the state for various exact, analytic calculations. First, we explicitly construct the MPS from the underlying spin-1/2 degrees of freedom. We then compute the transfer matrix of the state and obtain the normalization of the state. We then prove a result that is central to our paper: That $\ket{\psi_x}$ is a \textit{zero} eigenstate of $H$, for $h=0$. We finally compute various observables and two-point correlation functions, thereby characterizing the state. 
\begin{figure}[t]
    \includegraphics[width =  0.4 \textwidth, trim={0cm 0 0  0cm},clip]{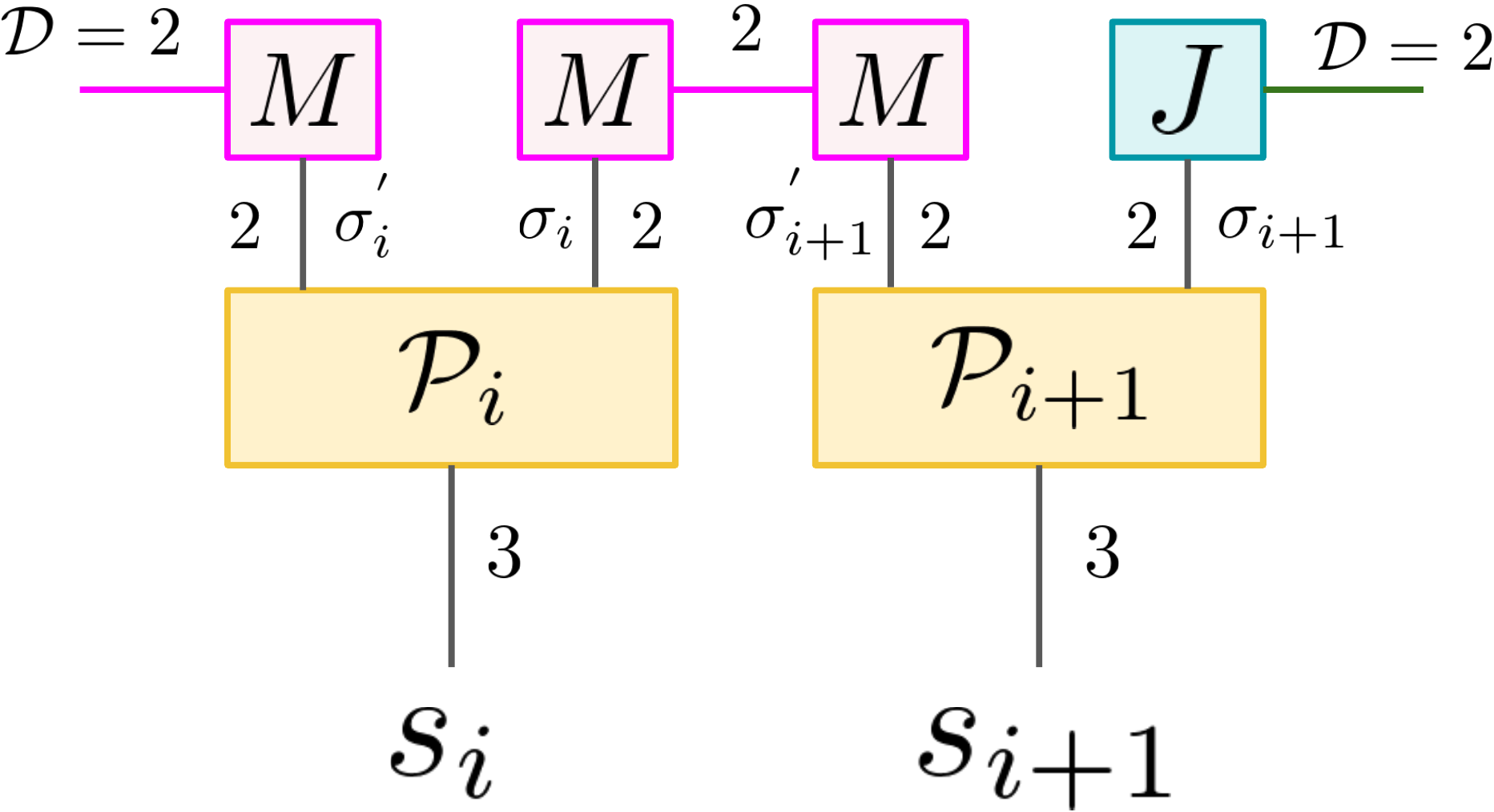}
\caption{Uncontracted MPS revealing the underlying spin-1/2 construction of $\ket{\psi_x}$. $M$, $J$ and $P$ are as given in Eqs.~\eqref{eqn:MJ} and ~\eqref{eqn:P}. The numbers denote the dimension of the index labelled. }
  \label{fig:MPSConstruction}
\end{figure}

\begin{figure*}[t]
    \includegraphics[width =  0.8 \textwidth]{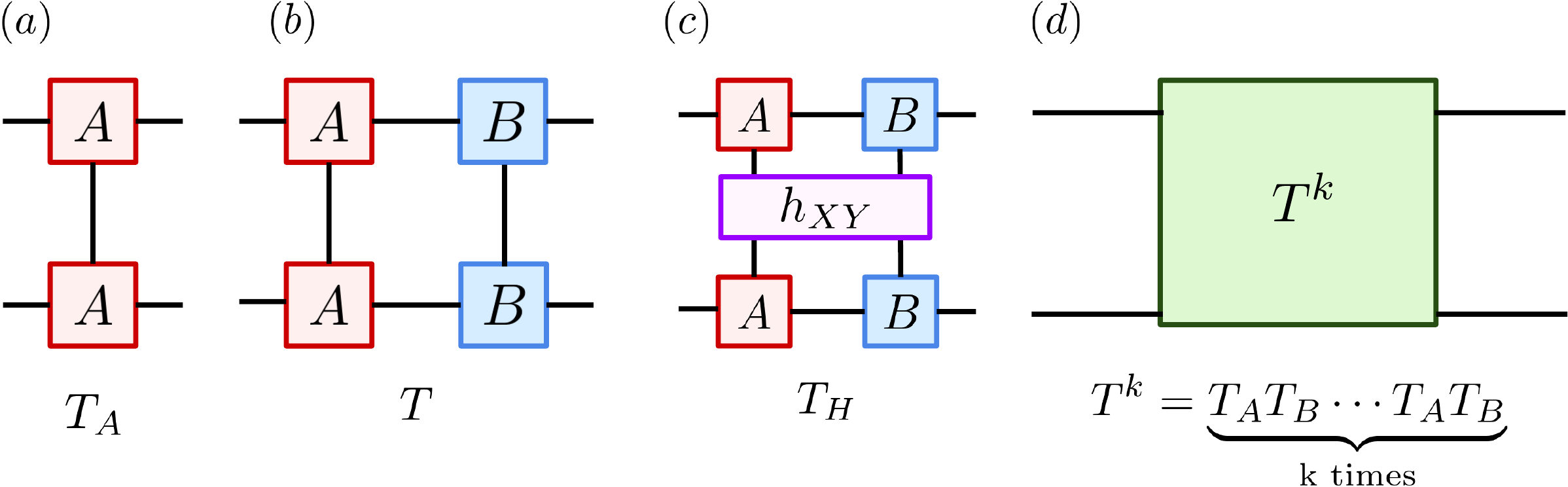}
\caption{Transfer operators. (a) Contractions showing the transfer matrix for the odd sites given by $A$, $T_A$, in the two-site unit cell translational invariant representation of $\ket{\psi_x}$. Note that one can identically construct $T_b$. (b) Two-site transfer matrix $T_{AB}$. Note that $T_{AB}=T_{BA}=T$. (c) Similarly, one can construct the transfer matrix $T_H$ for the two-body term $H_{i,i+1}^{XY} =S^x_i S^x_{i+1} + S^y_i S^y_{i+1}$ (in future abbreviated as $h_{XY})$. $\langle h_{XY} \rangle$ is found by tracing over the dangling bond dimensions. (d) $T^k$, the two-site matrix taken to the k\textsuperscript{th} power.}
  \label{fig:TransferMatrices}
\end{figure*}

\subsection{Constructing the MPS State}

We first construct the state on $L$ sites from underlying $2L$ spin-1/2 degrees of freedom. As discussed in the main text, $\ket{\psi_x}$ can be constructed by first laying down $\ket{o} = \frac{1}{\sqrt{2}} (\ket{\upuparrows}+\ket{\downdownarrows})$ on sites $(4i, 4i+1)$ and $\ket{g} = \frac{1}{\sqrt{2}} (\ket{\upuparrows}-\ket{\downdownarrows})$ on sites $(4i+2,4i+3)$ and then `projecting' pairs $(2i-1,2i)$ of spin-1/2s onto spin-$1$ degrees of freedom at site $i$ on the \textit{physical} chain using the local map $\mathcal{P}_i = \ket{1}_i\bra{\upuparrows}_{2i-1, 2i}+ \ket{0}_i(\ket{ \updownarrows}_{2i-1, 2i} +\ket{\downuparrows}_{2i-1, 2i})+\ket{-1}_i\bra{\downdownarrows}_{2i-1, 2i}$. This AKLT-like construction naturally yields an MPS representation for $\ket{\psi_x}$.

In order to do so, we first find the MPS representations of $\ket{o}$ and $\ket{g}$, which are simply: $\ket{o} = \sum_{\sigma,\sigma'} (MM)_{\sigma, \sigma'} \ket{\sigma, \sigma'}$ and $\ket{g} = \sum_{\sigma,\sigma'} (MJ)_{\sigma, \sigma'} \ket{\sigma, \sigma'}$, where $M$ and $J$ are given by:
\begin{align}
\label{eqn:MJ}
\begin{tabular}{ccc}
$
M = \frac{1}{2^{1/4}} \begin{pmatrix}
1&0\\
0 & 1\\
\end{pmatrix}$; &$
J = \frac{1}{2^{1/4}}\begin{pmatrix}
1&0\\
0& -1\\
\end{pmatrix}$ 
\end{tabular},
\end{align} 
where $\sigma,\sigma'=1,2$ and $|1\rangle = |\uparrow\rangle, |2\rangle = |\downarrow \rangle$.
We then  apply the local map $\mathcal{P}_i$, which maps spin-1/2s from sites $2i-1$ and $2i$ onto a single spin-1 at site $i$. Note that $P_i$,   a 4-by-3 matrix, can be reshaped into a 2-3-2 tensor and expressed in the following MPS conducive manner:
\begin{align}
\label{eqn:P}
\begin{tabular}{ccc}
$
(P_{i})^{-1} = \begin{pmatrix}
1&0\\
0 & 0\\
\end{pmatrix}$; &$
(P_{i})^{0} = \begin{pmatrix}
0&1\\
1 & 0\\
\end{pmatrix}$; &  $
( P_{i})^{1} = \begin{pmatrix}
0&0\\
0 & 1 \\
\end{pmatrix}$ 
\end{tabular}.
\end{align}

We finish the construction of the MPS representation by contracting $M$, $P$, and $M$ to form $A$ (given in the main text) and $M$, $P$, and $J$ to form $B$, as follows (see Fig.~\ref{fig:MPSConstruction}):
\begin{align*}
A_{\sigma,\sigma'}^{s} = \sum_{\rho, \rho'} M_{\sigma,  \rho}(P_i)_{\rho,\rho'}^{s}M_{\rho',\sigma'} \\
B_{\sigma, \sigma'}^{s} = \sum_{\rho, \rho'} M_{\sigma,  \rho}(P_i)_{\rho,\rho'}^{s}J_{\rho',\sigma'}, 
\end{align*}
 where $\sigma,\sigma',\rho,\rho' = 1,2$ are bond-indices with $s=-1,0,1$ is the physical index (mapping to the local spin-1 states $|s\rangle \in \{ |-1\rangle, |0\rangle, |1\rangle \}$).

\begin{figure*}[t] 
    \includegraphics[width =  0.95 \textwidth, trim={0cm 0 0  0cm},clip]{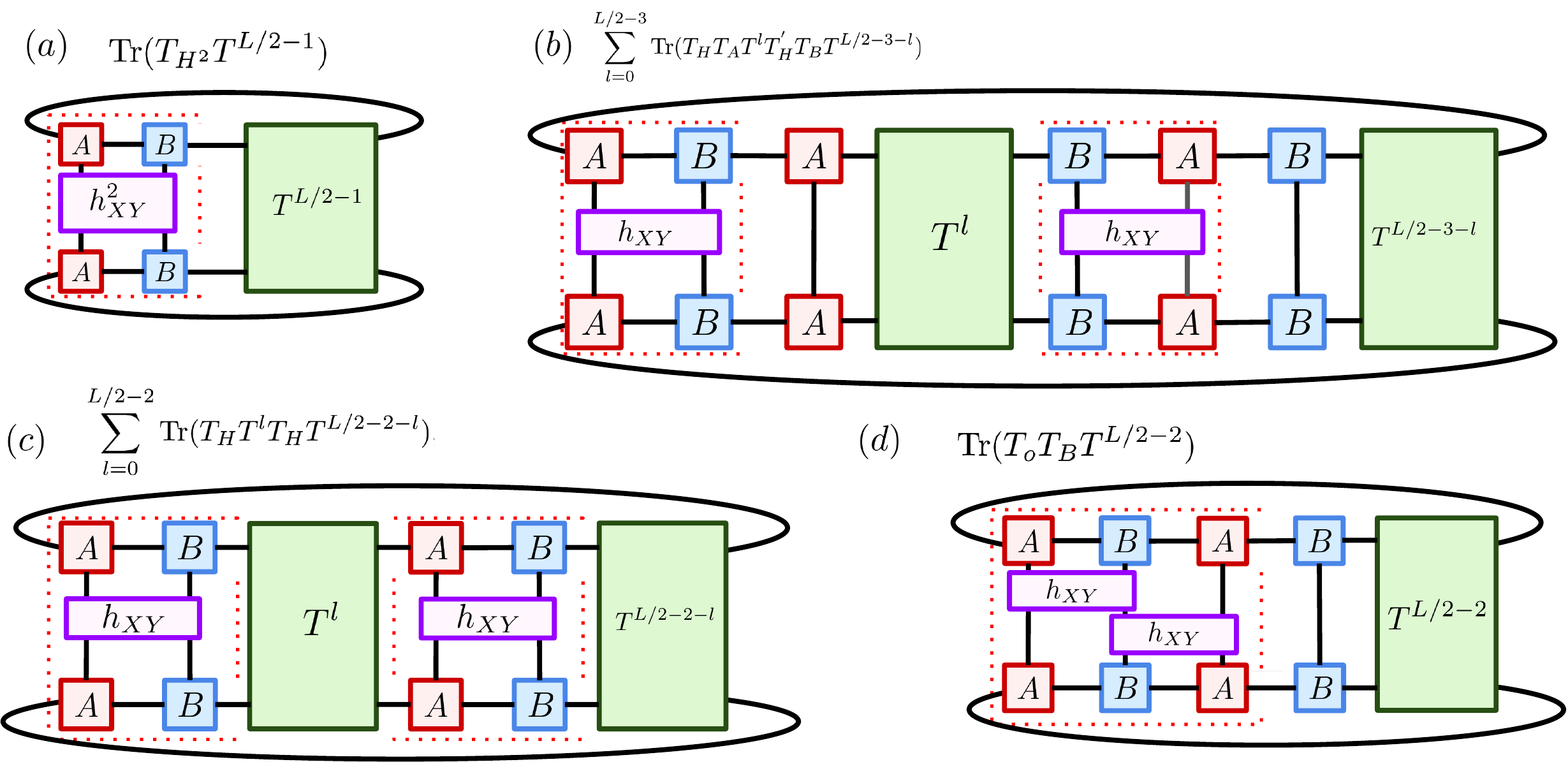}
\caption{Contractions for the calculation of $\langle H_{\text{XY}}^2 \rangle$. The various contractions are produced by placing one two-body term $h_{\text{XY}}$ first on the first two sites, and then placing the second two-body term somewhere along the lattice. 
(Due to two-site translational invariance, this covers half the contributions of the terms in the double sum  of $H^2_{\text{XY}}$. The other half, consists of shifting all diagrams by one site.)
The green tensors $T^k$ are transfer matrices raised to power $k$, introduced in fig.~\ref{fig:TransferMatrices}(d).(a) ``Diagonal term" (see Eq. \ref{eq:t1}) where both $h_{\text{XY}}$s are placed on the first two sites. Red dotted box indicate the contraction which gives $T_{H^2}$. (b) A term in which the second $h_{\text{XY}}$ is placed on ``even-odd" sites, in that order (Eq. \ref{eq:t4}). The two red dotted boxes show $T_{H}$ and $T_H^{'}$, respectively. (c) A term similar to (b), but where $h_{\text{XY}}$'s are placed along only ``odd-even" sites ((Eq. \ref{eq:t3}). (d) ``Overlap" term (Eq. \ref{eq:t2}) in which the second $h_{\text{XY}}$ is placed one site to the  right of the first one. Note that an identical term (\ref{eq:t5}) exists where $h_{\text{XY}}$s are placed on the left. The red dotted box gives $T_O$, $T_O'$, respectively. Together, these terms cancel. } 
  \label{fig:Contractions}
\end{figure*}

\subsection{Transfer Matrices and Normalizing $\ket{\psi_x}$ }
 \label{App:B}

In this section, we compute the single and two-site transfer matrices. As shown in Fig.~\ref{fig:TransferMatrices}(a), the single site transfer matrix is simply found by contracting together the middle legs of two 2-3-2 tensor $A$. The object now obtained is a 2-2-2-2 tensor. The transfer \textit{matrix}, a 4-by-4 tensor, is found by transposing the middle two legs and reshaping the 2-2-2-2 tensor into a 4-by-4 matrix. Similarly, $T_B$ can be found by contracting in an identical manner a pair of $B$s, and then transposing, and reshaping as described previously. Simiarly, the two-site transfer matrices can be found,
$T_{AB} = T_{BA} = T_A T_B = T_B T_A = T$, see Fig.~\ref{fig:TransferMatrices}(b). We have that
\begin{align*} 
&T_A = \begin{pmatrix}
\frac{1}{2} &0 & 0 & \frac{1}{2}\\
0 & 0 & \frac{1}{2} & 0 \\
0 & \frac{1}{2} & 0 & 0 \\
\frac{1}{2} &0 & 0 & \frac{1}{2}
\end{pmatrix};   
T_B = \begin{pmatrix}
\frac{1}{2} &0 & 0 & \frac{1}{2}\\
0 & 0 & -\frac{1}{2} & 0 \\
0 & -\frac{1}{2} & 0 & 0 \\
\frac{1}{2} &0 & 0 & \frac{1}{2}
\end{pmatrix};   \nonumber \\
&T=T_{AB} = \begin{pmatrix}
\frac{1}{2} &0 & 0 & \frac{1}{2}\\
0 & -\frac{1}{4} & 0 & 0 \\
0 & 0 & -\frac{1}{4} & 0 \\
\frac{1}{2} &0 & 0 & \frac{1}{2}
\end{pmatrix}.
\end{align*}

As $T$ is Hermitian, it has the same right and left eigenvectors $|r_i)$ and $|l_i)$, with eigenvalues $\gamma_i$ being $1, -\frac{1}{4}, -\frac{1}{4}$ and $0$.
In particular, the dominant eigenvector is found to be given by $\frac{1}{\sqrt{2}} (1,0,0,1)^\dagger$. 
This property will be an important fact for the calculating of the reduced density matrices later.
Note that the existence of a single dominant eigenvalue of $1$ already immediately implies that the correlation functions are exponentially decaying and that the state is normalized in the TDL. Furthermore, it also implies that the state obeys the cluster decomposition theorem $\lim_{|x-y| \to \infty} \langle O_x O_y \rangle - \langle O_x \rangle \langle O_y \rangle = 0$ for local $O_x, O_y$, signifying that it is a `physical' state.

From the eigenvalues of the transfer matrix it is evident that the normalization is given explicitly by: $\mathcal{N}^2 = \braket{\psi_x}{\psi_x} = \textrm{Tr}(T)^{L/2} = 1+2(-\frac{1}{4})^{L/2}$. Note that this deviation from unit normalization for any finite $L$ can be understood from the fact that the middle state $\ket{\psi_{L/2}}$ in the tower of scarred states $|\psi_n\rangle$ (which make up $|\psi_x\rangle$,   is not normalized, when constructed by applying the map $\mathcal{P} = \bigotimes_i \mathcal{P}_i$ onto a normalized $|\phi_n\rangle$ (as explained in the main text). It is rather simple to see why this occurs.

To generate the middle state $|\psi_{L/2}\rangle$ in the tower, recall we can start from the state $|\phi_{L/2}\rangle$ at the spin-1/2 level. For the latter, the ladder operator $J^+ = \sum_{i=1}^L (-1)^i s^+_{2i} s^+_{2i+1}$ must be applied $L/2$ times from the lowest-weight state of $J^z = \sum_i^{2L} s^z_i$ in the largest spin-representation of $\vec{J}\cdot\vec{J}$, which is the state $|\downarrow \downarrow \downarrow \downarrow \cdots\rangle$.
By doing so, a superposition of ${L \choose L/2}$ different orthonormal product states in the $z$-basis ($|\uparrow \downarrow \downarrow \uparrow \cdots \rangle$ etc.) is generated, with  coefficients of equal magnitude albeit differing signs due to the definition of the ladder operator.  
At this stage, $|\phi_n\rangle$ can be made normalized, $|\phi_n\rangle \mapsto \frac{1}{\sqrt{\mathcal N}} |\phi_n\rangle$: the superposition of all the product states which make it up gives the normalization constant $\mathcal{N} = {L \choose L/2}$. 

Now consider what happens upon applying the map $\mathcal{P}$. All spin-1/2 product states that make up $|\phi_{L/2}\rangle$ map uniquely to a counterpart spin-1  product state, except for two product states, $\ket{\uparrow \downarrow \downarrow \uparrow \dots}$ and $\ket{\downarrow \uparrow  \uparrow \downarrow \dots}$. These instead both map to $\ket{\textbf{0}} = \ket{000 \dots}$. For $L = 4n$, these states come with the same sign and constructively interfere, mapping to therefore give 2 copies of $\ket{\textbf{0}}$. For $L=4n+2$ ($n>0$), the states have different signs at the spin-1/2 level, and therefore destructively interfere and do not give any contribution when mapped onto the spin-1 level. Thus, for $L=4n$, in the superposition of the spin-1 product states making up $|\psi_{n=L/2}\rangle$, there are ${L \choose L/2}-2$ orthonormal product states with equal magnitude contribution and a copy of $\ket{\textbf{0}}$ entering with twice the magnitude of the other terms. The norm of the state $|\psi_{L/2}\rangle$ is therefore $\sqrt{\frac{{L \choose L/2}+2}{{L \choose L/2}}}$. On ther other hand, for $L=4n+2$, there are ${L \choose L/2}-2$ orthonormal states with equal magnitude and no copies of $\ket{\textbf{0}}$, giving a normalization of $\sqrt{\frac{{L \choose L/2}-2}{{L \choose L/2}}}$. 

It is possible to see, from similar considerations, that for $n\neq L/2$ this constructive/destructive interference does not happen --- the $|\psi_n\rangle$ constructed by mapping from normalized $|\phi_n\rangle$ are all normalized.
Thus, the deviation from unity of the normalization of the MPS can be understood as arising from a deviation from unity of the normalization of $|\psi_{L/2}\rangle$.

\subsection{$\ket{\psi_x}$ is an eigenstate of spin-$1$ $XY$ model}

In this section we prove that $H \ket{\psi_x}=0$ for $h=0$. We show this by calculating $\langle H \rangle$ and $\langle H^2 \rangle$ using the MPS representation and showing that both are $0$, thus proving that $H \ket{\psi_x}=0$. Note that the calculation of the latter quantity $\langle H^2\rangle$ is actually sufficient to show this result due to the positive definiteness  of the inner product. We first calculate the single site energy expectation, then show that $\langle H_{\text{XY}}^2 \rangle$=0, and finally show that  $\langle V^2 \rangle$=0.

The energy calculation is straightforward : $\sum_{j=1}^4 \gamma_j^{L/2-1} (l_j | T_H | r_j )=0$, where $\gamma_j, |l_j), |r_j)$ are the eigenvalues, left and right eigenvectors of the transfer matrix $T$ and $T_H$ is the local energy operator sandwiched between matrices $A,B$, as drawn in Fig.~\ref{fig:TransferMatrices}(b). Such a result  is also true upon swapping $B \leftrightarrow A$. Thus, $\langle H \rangle = 0$ in total.

We now calculate $\langle H_{\text{XY}}^2 \rangle$. Note that $h_{\text{XY}}$ is used as short-hand (with self-evident context), $h_{\text{XY}} = H_{i,i+1}^{\text{XY}}$. We expand $H_{\text{XY}}^2 = \sum_{i=1}^{L} \sum_{j=1}^{L} H_{i, i+1}^{\text{XY}} H_{j, j+1}^{\text{XY}}$. This leads to four classes of terms as shown in Fig. \ref{fig:Contractions}:
(a) ``diagonal" terms where two $h_{\text{XY}}$ act on the same pair of sites, %
(b) terms for which two $h_{\text{XY}}$ s do not overlap and begin on sites of different parities (for example, one $h_{\text{XY}}$ is placed on sites (1,2),another placed on sites (5, 6)),
(c) terms for which two  $h_{\text{XY}}$s do not overlap and begin on sites with the same parity (for example, one $h_{\text{XY}}$ is placed on sites (1,2),another placed on sites (3, 4)),
and (d) ``overlap" terms for which two $h_{\text{XY}}$ s straddle three sites (for example fixing a particular pair of sites $(1,2)$   where the first $h_{\text{XY}}$ acts, we place the second $h_{\text{XY}}$ one site to the left and right of it, so there are actually two such terms, on sites $(2,3)$ and $(L,1)$). 

\begin{figure}[t]
    \includegraphics[width =  0.4\textwidth]{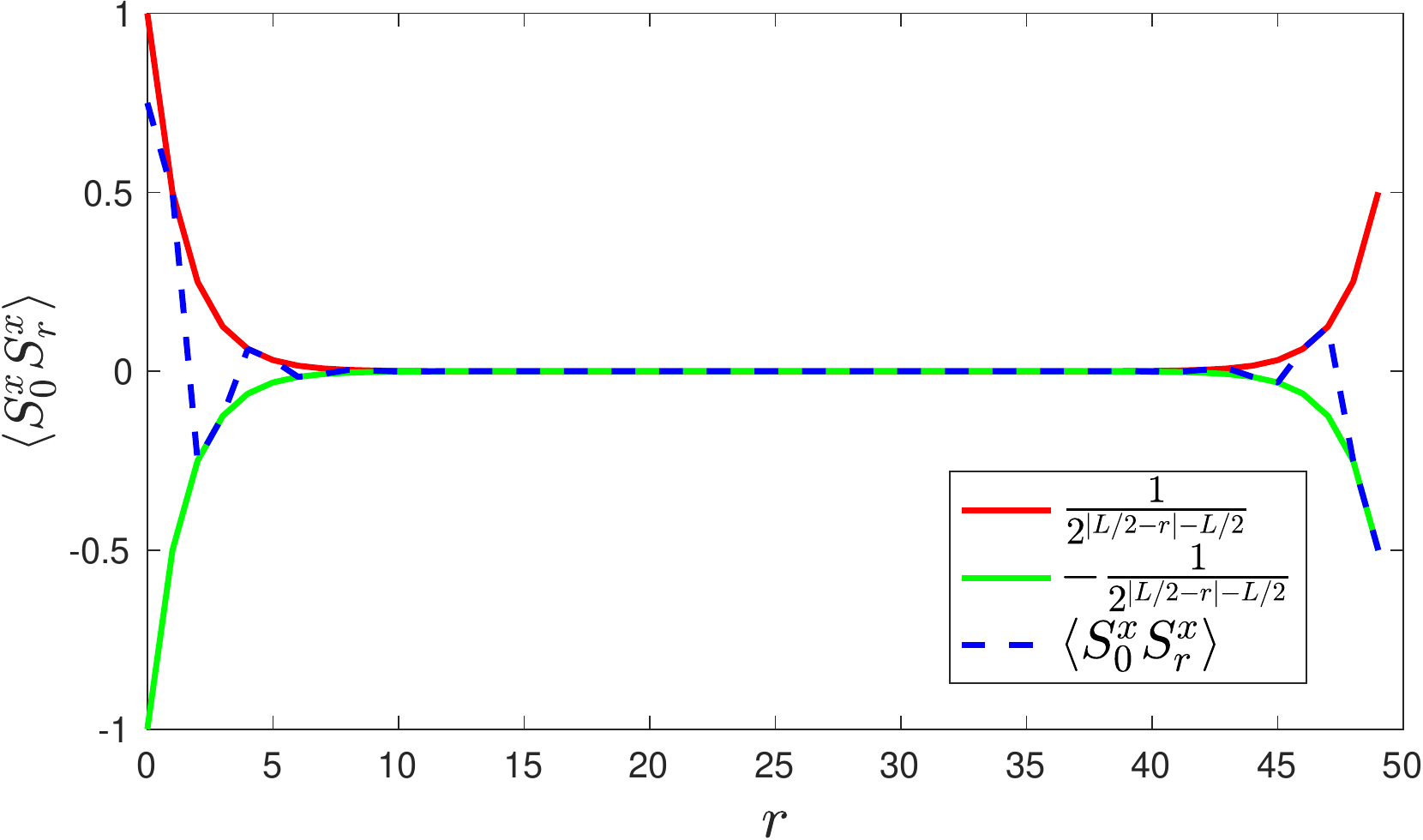}
\caption{Two point correlation function $\langle S_0^{x} S_r^{x} \rangle$ plotted as a function of $r$, the distance along the ring from the origin, for $L=50$. Note that the correlation function seems to decay as $\frac{1}{2^{l}}$, where $l$ is the absolute distance from the origin.
}
  \label{fig:Correlations}
\end{figure}

It is possible to exploit the two-site translational invariance of the MPS representation by selecting, once for each parity, the site at which the first $h_{XY}$ begins, and then placing a second $h_{XY}$ anywhere on the lattice. However we find that there is no difference between placing the first $h_{XY}$ beginning at an odd site (where tensor $A$ acts) or an even site (where tensor $B$ acts). The MPS contractions (we only show it for $A$) give the following: 
\begin{align}
\label{eq:t1}
& \textrm{Tr}(T_{H^2}T^{L/2-1}) = 1+(-1)^{L/2}2^{2-L} \\
\label{eq:t2}
& \textrm{Tr}(T_{o}T_{B}T^{L/2-2})=-\frac{1}{2}-3(-1)^{L/2}2^{1-L}
\\
\label{eq:t5}
& \textrm{Tr}(T_{o}^{'}T^{L/2-2}T_{A}) =-\frac{1}{2}-3(-1)^{L/2}2^{1-L}
%
\\
\label{eq:t3}
& \sum_{l=0}^{L/2-2}\textrm{Tr}(T_{H}T^{l}T_{H}T^{L/2-2-l})  = (L-2)(-1)^{L/2}2^{2-L} 
\\
\label{eq:t4}
& \sum_{l=0}^{L/2-3}\textrm{Tr}(T_{H}T_{A}T^{l}T_{H}^{'} T_{B} T^{L/2-3-l}) = (L-4)(-1)^{L/2-1}2^{2-L},
\end{align}
where $T_{O}$ is the transfer operator which captures the situation in which two $h_{XY}$s straddled the first three terms, while $T_{O}'$ is the transfer operator which straddles the last term and the first two terms. Note that $T_{O} T_B = T_{O}^{'} T_A$. Taken together, all terms sum to 0 for any $L$ and thus, $\langle H_{\text{XY}}^2 \rangle =0$, implying that $H_{\text{XY}} \ket{\psi_x} = 0$. One can go through the same calculations (with the same contractions), swapping out $h_{XY}$ for $V_i = (S^{+}_i)^2(S^{-}_{i+1})^2 + (S^{-}_i)^2(S^{+}_{i+1})^2$. In this case however, all of the terms which are analogous to the above terms in Eq. \eqref{eq:t1}-\eqref{eq:t4} are identically 0. 
(It is actually easy to see $V_i | \psi_x\rangle = 0$ through different means because $V_i = |-1,1\rangle \langle 1,-1| + \text{h.c.}$ and in the expansion of $|\psi_n\rangle$ over product states in the $z$-basis, there are never any local $|-1,1\rangle$ or $|1,-1\rangle$ configurations.)

Thus, all together, these calculations show that $(H_\text{XY} + V) \ket{\psi_x} = 0$.

\subsection{Observables and Correlation Functions}

Using the transfer matrices and operators defined above, it is straightforward to compute single-site spin observables and two-point spin-spin correlation functions for the state $\ket{\psi_x}$. Note that the calculations are very similar to the calculations shown above. First, we note that $\langle S_i^z \rangle = \langle S_i^y \rangle = \langle S_i^x \rangle =0$, at any site $i$. We now present the two-point (connected) correlation functions. We note that $\langle S_i^z S_j^z \rangle = 0$ if $|i-j|>1$: For $|i-j|=1$, $\langle S_i^z S_j^z \rangle =\frac{1}{4}$ and for $i=j$, $\langle (S_i^z)^2 \rangle = \frac{1}{2}$. Now, we present the $S^x$ and $S^y$ correlation functions.

For $i$ odd and $j$ even: 
\begin{align}
  & \langle S_i^{y} S_{i+r}^{y} \rangle_c =\langle S_j^{x} S_{j+r}^{x} \rangle_c \nonumber \\
  & =
  \begin{cases}
                                   \frac{3}{4}+(-1)^{L/2}2^{1-L}& r=0 \\
                                  (-1)^{\frac{r+1}{2}}(\frac{1}{2^{r}}+(-1)^{L/2}\frac{1}{2^{L-r}}) & r \text{ odd} \\
  (-1)^{\frac{r}{2}}(\frac{1}{2^{r}}+(-1)^{L/2}\frac{1}{2^{L-r}}) & r \ne 0, \text{even} 
  \end{cases}
\end{align}

For $i$ odd and $j$ even: 
\begin{align}
  & \langle S_i^{x} S_{i+r}^{x} \rangle_c =\langle S_j^{y} S_{j+r}^{y} \rangle_c \nonumber \\
  & =
  \begin{cases}
                                   \frac{3}{4}+(-1)^{L/2}2^{1-L}& r=0 \\
                                  (-1)^{\frac{r-1}{2}}(\frac{1}{2^{r}}+(-1)^{L/2}\frac{1}{2^{L-r}}) & r \text{ odd} \\
  (-1)^{\frac{r}{2}}(\frac{1}{2^{r}}+(-1)^{L/2}\frac{1}{2^{L-r}}) & r > 0, \text{even} 
  \end{cases}
\end{align}

An example of one of these two-point correlation functions is provided in Fig. \ref{fig:Correlations}.

\section{Entanglement Entropy of $\ket{\psi_x}$}
\label{Appendix:EE}

\begin{figure*}[t]
    \includegraphics[width =  0.7 \textwidth]{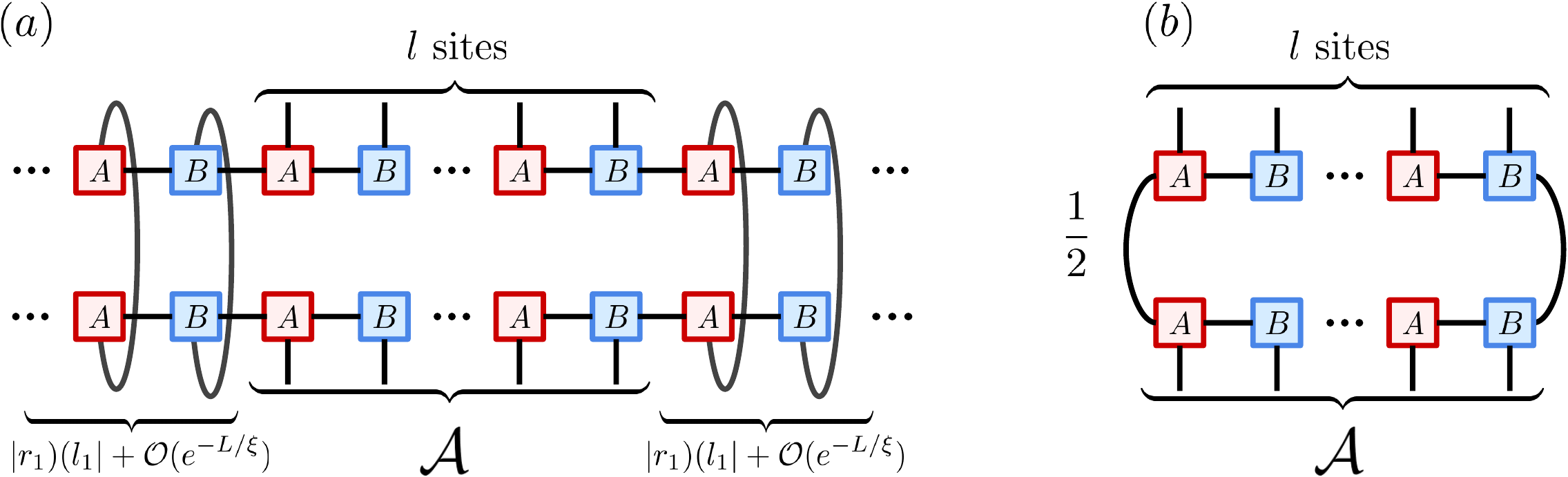}
\caption{The density matrix of a particular subregion $\mathcal{A}$, $\rho_{\mathcal{A}}$ in the TDL. (a) The exact density matrix. For large $L$, the transfer matrices on the left and right can simply be replaced by $|r_1)(l_1|$, the projector onto the dominant eigenvector, with corrections that are exponentially small in $L$. (b) The  density matrix in the thermodynamic limit for sub-region $\mathcal{A}$ which has $l$ sites.}
  \label{fig:RDM}
\end{figure*}

\begin{figure*}[t]
    \includegraphics[width =  0.8 \textwidth]{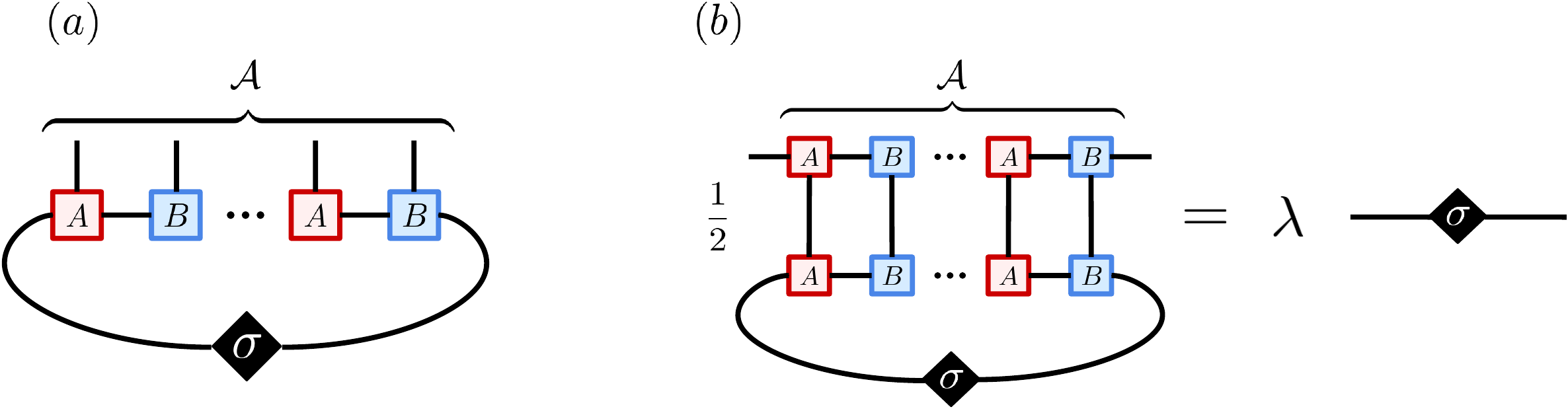}
\caption{Finding the eigenvectors of $\rho_{\mathcal{A}}$. (a) $|\Lambda \rangle$, ansatz for an eigenvector of $\rho_\mathcal{A}$. (b) A sufficient equation for $|\Lambda \rangle$ to be an eigenvector of $\rho_\mathcal{A}$.}
  \label{fig:eigenequations}
\end{figure*}

\subsection{Eigenvalues of the reduced density matrix of $|\psi_x\rangle$}

We prove here the result that the reduced density matrix $\rho_\mathcal{A}$ of $|\psi_x\rangle$ (in the TDL), for a (contiguous) region $\mathcal{A}$ comprised of sites $1,\cdots, l$, has eigenvalues
\begin{align}
\lambda_1 = \frac{1}{4}, \lambda_2 = \frac{1}{4}, \lambda_3 = \frac{1}{4} + \frac{1}{2^{l+1}}, \lambda_4 = \frac{1}{4} - \frac{1}{2^{l+1}}. 
\end{align} 
Our proof will also furnish the four eigenvectors $|\Lambda_i\rangle$, $i = 1,\cdots,4$ of $\rho_A$. The reduced density matrix, is given in Fig.~\ref{fig:RDM}(a).

We start by noticing that the transfer matrices $T_A, T_B, T = T_A T_B$ of the matrices $A,B$ making up the MPS $|\psi_x\rangle$, have a single dominant left $( l_1 | $ and right eigenvector  $| r_1 )$ with unit eigenvalue.
In fact, 
\begin{align}
| l_1 ) = | r_1 ) = \frac{1}{\sqrt{2}} (1,0,0,1)^\dagger,
\end{align}
which we note is simply the identity matrix $\begin{pmatrix} 1 & 0 \\ 0 & 1 \end{pmatrix}$ (a $2-2$ tensor) reshaped into a vector (a $4 - 1$ tensor), multiplied by the coefficient $1/\sqrt{2}$.

Thus, in the TDL, $\rho_{\mathcal{A}}$ can simply be written as  
\begin{align}
\rho_\mathcal{A} = \frac{1}{2} \left( l_1  \Bigg|  \prod_{i=1}^l   O_i \otimes O_i^\dagger       \Bigg| r_1 \right),
\end{align}
where
\begin{align}
O_i = \begin{cases}
& A_{-1} |-1\rangle_i + A_0 |0\rangle_i + A_1 |1\rangle_i, \text{ for } i \text{ odd}, \nonumber \\
& B_{-1} |-1\rangle_i + B_0 |0\rangle_i + B_1 |1\rangle_i, \text{ for } i \text{ even}, \nonumber \\
\end{cases}
\end{align}
see Fig. \ref{fig:RDM}(b). 
Note the   vectors act on different spaces: $|l_1), |r_1)$ are vectors on the bond indices of $O_i \otimes O_i^\dagger$, while $|-1\rangle, |0\rangle, |1\rangle$ correspond to vectors on the physical indices.

\begin{figure}[t]
    \includegraphics[width =  0.5 \textwidth, trim={0cm 0 0  0cm},clip]{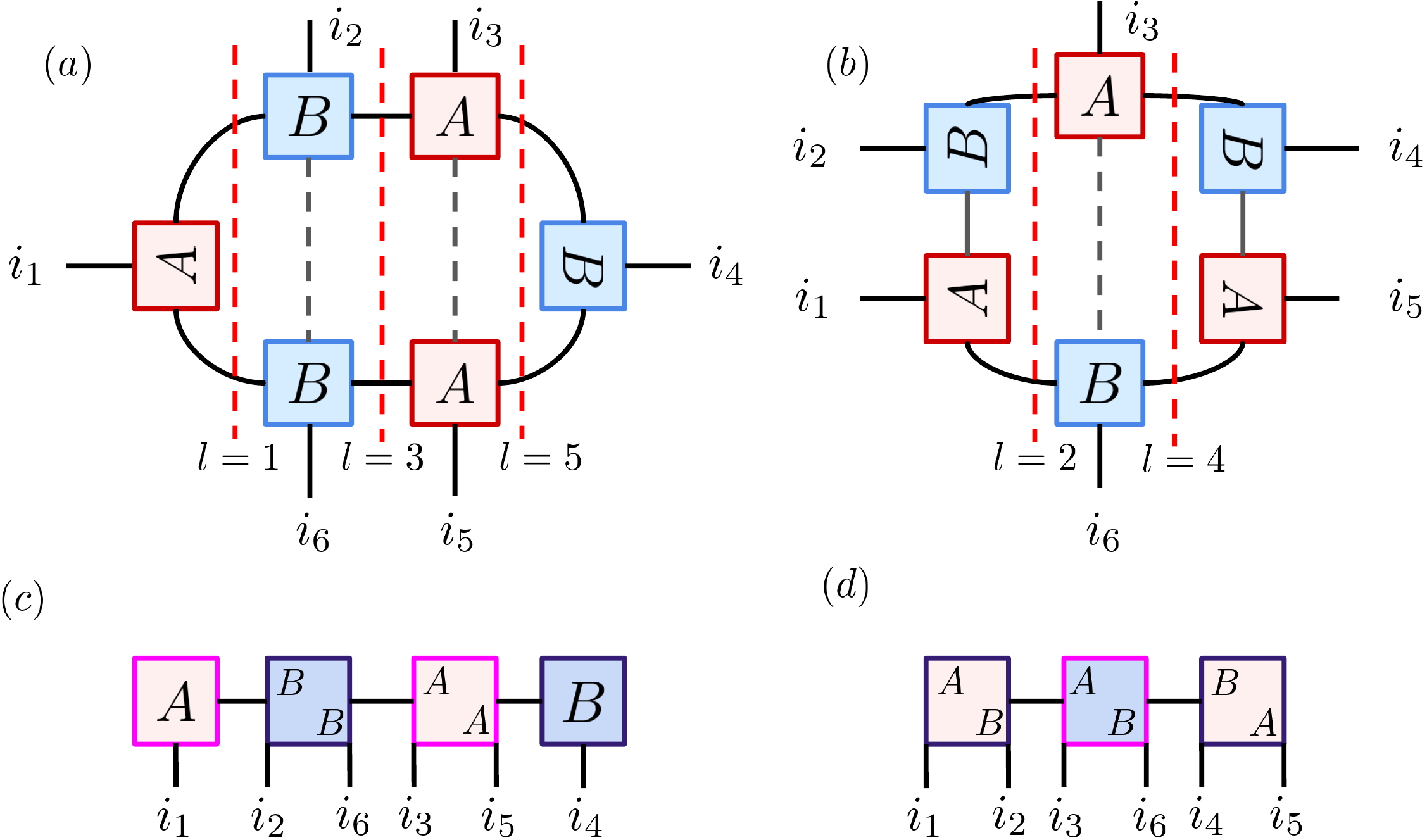}
\caption{(a,c) (above) Annotated MPS (with PBC) representation of $\ket{\psi_x}$ for $L=6$ and (below) corresponding OBC MPS representation. When converting the PBC MPS representation of $\ket{\psi_x}$ to an OBC MPS representation, the following procedure is taken. The first (middle) site is left alone as a 1-9-4 (4-9-1) tensor, while the other, originally 2-3-2, tensors are grouped together as 4-3-3-4 tensors. The black dashed lines refer to ``pairings" of the MPS tensors, made across the ``equator" of the PBC MPS. The entanglement cuts are illustrated by the red dashed lines. (b,d) Corresponding illustrations for the OBC MPS representation which gives the Schmidt decomposition which allows us to calculate entropies for cuts made at even sites. Note that this representation begins with the first two sites paired together as a 4-3-3-1 tensor. The rest proceeds similarly to the odd case.}
  \label{fig:MPSDoubled}
\end{figure}

We consider as an ansatz
\begin{align}
| \Lambda \rangle = \sum_{s_1, \cdots s_l} \text{Tr} \left( A_{s_1} B_{s_2} A_{s_3} \cdots  \sigma \right) |s_1, \cdots s_l\rangle,
\end{align}
 to be an eigenvector of $\rho_\mathcal{A}$, 
which is nothing but the original MPS but on the subregion $\mathcal{A}$, with appropriate boundary conditions given by the matrix $\sigma$, see Fig.~\ref{fig:eigenequations}(a). We parameterize this as
\begin{align}
\sigma = \begin{pmatrix}
a & b \\ 
c & d
\end{pmatrix}.
\end{align}
The condition to be solved then reads $|\Lambda \rangle$ is an eigenvector of $\rho_\mathcal{A}$, that is, $\rho_\mathcal{A} | \Lambda \rangle = \lambda |\Lambda \rangle$.  
It is sufficient (though not necessary) to solve the  expression, written in diagrams in Fig.~\ref{fig:eigenequations}(b), for $|\Lambda \rangle$.

Consider first even $l$. Upon evaluating Fig.~\ref{fig:eigenequations}(b), we have that
\begin{align}
\begin{pmatrix}
\frac{a}{4} +  (-1)^{\frac{l}{2}} \frac{1}{2^{l+1}} d & \frac{b}{4} \\ 
\frac{c}{4} & \frac{d}{4} +  (-1)^{\frac{l}{2}} \frac{1}{2^{l+1}} a
\end{pmatrix}
 = \lambda \begin{pmatrix}
a & b \\ 
c & d
\end{pmatrix}.
\end{align}
Thus we have the solutions
\begin{align}
& \lambda_1 = \frac{1}{4}, \qquad \qquad  \sigma_1 = 
\begin{pmatrix}
0 & 1 \\ 
0 & 0
\end{pmatrix}; \nonumber \\
& \lambda_2 = \frac{1}{4}, \qquad \qquad 
\sigma_2 = 
\begin{pmatrix}
0 & 0 \\ 
1 & 0
\end{pmatrix}; \nonumber \\
& \lambda_3 = \frac{1}{4} + \frac{1}{2^{l+1}},
~~ \sigma_3 = 
\begin{pmatrix}
1 & 0 \\ 
0 & (-1)^{l/2}
\end{pmatrix}; \nonumber \\
& \lambda_4 = \frac{1}{4} -  \frac{1}{2^{l+1}},
~~ \sigma_4 = 
\begin{pmatrix}
1 & 0 \\ 
0 & (-1)^{l/2+1}
\end{pmatrix},
\end{align}
with corresponding $|\Lambda_i\rangle$. We also have to check that $| \Lambda_i\rangle$ are orthogonal. We   compute the overlap matrix
\begin{align}
M_{ij} = \text{Tr}\left( \prod_{n=1}^l T_{O_n} (\sigma_i^* \otimes \sigma_j) \right),
\end{align}
where $T_{O_n}$ is the transfer matrix equal to $T_A$($T_B$) for odd(even) sites. We get
\begin{align}
M_{ij} = \begin{pmatrix}
\frac{1}{2} & 0 & 0 & 0 \\
0 & \frac{1}{2} & 0 &  0 \\
0 & 0 & 1 + 2^{1-l} & 0 \\
0 & 0 & 0 & 1 - 2^{1-l}
\end{pmatrix}.
\end{align}
Thus, we see that $| \Lambda_i \rangle$, $i=1,\cdots,4$ are orthogonal (and in particular $|\Lambda_1\rangle \neq |\Lambda_2\rangle$), and that the eigenvalues sum to 1 (so $\Tr(\rho_\mathcal{A}) = 1$). Therefore $\rho_\mathcal{A}$ has eigenvalues $(1/4, 1/4, 1/4+1/2^{l+1}, 1/4-1/2^{l+1})$.

Next consider odd $l$. Upon evaluating Fig.~\ref{fig:eigenequations}(b), we have that
\begin{align}
\begin{pmatrix}
\frac{a}{4}  & \frac{b}{4} + (-1)^{\frac{l-1}{2}} \frac{1}{2^{l+1}} c \\ 
\frac{c}{4} + (-1)^{\frac{l-1}{2}} \frac{1}{2^{l+1}} b & \frac{d}{4}
\end{pmatrix}
 = \lambda \begin{pmatrix}
a & b \\ 
c & d
\end{pmatrix}.
\end{align}
Thus we have the solutions
\begin{align}
& \lambda_1 = \frac{1}{4}, \qquad \qquad  \sigma_1 = 
\begin{pmatrix}
1 & 0 \\ 
0 & 0
\end{pmatrix}; \nonumber \\
& \lambda_2 = \frac{1}{4}, \qquad \qquad 
\sigma_2 = 
\begin{pmatrix}
0 & 0 \\ 
0 & 1
\end{pmatrix}; \nonumber \\
& \lambda_3 = \frac{1}{4} + \frac{1}{2^{l+1}},
~~ \sigma_3 = 
\begin{pmatrix}
0 & 1 \\ 
(-1)^{(l-1)/2} & 0
\end{pmatrix}; \nonumber \\
& \lambda_4 = \frac{1}{4} -  \frac{1}{2^{l+1}},
~~ \sigma_4 = 
\begin{pmatrix}
0 & 1 \\ 
(-1)^{(l+1)/2} & 0
\end{pmatrix},
\end{align}
with corresponding $|\Lambda_i\rangle$.   We   compute the overlap matrix and get
\begin{align}
M_{ij} = \begin{pmatrix}
\frac{1}{2} & 0 & 0 & 0 \\
0 & \frac{1}{2} & 0 &  0 \\
0 & 0 & 1 + 2^{1-l} & 0 \\
0 & 0 & 0 & 1 - 2^{1-l}
\end{pmatrix}.
\end{align}
Thus, (aside from the case $l=1$, but this is easily handled) we see that $| \Lambda_i \rangle$, $i=1,\cdots,4$ are orthogonal and that the eigenvalues sum to 1, and therefore $\rho_\mathcal{A}$ has eigenvalues $(1/4, 1/4, 1/4+1/2^{l+1}, 1/4-1/2^{l+1})$, as claimed.

\begin{figure}[t]
    \includegraphics[width =  0.4 \textwidth, trim={0cm 0 0  0cm},clip]{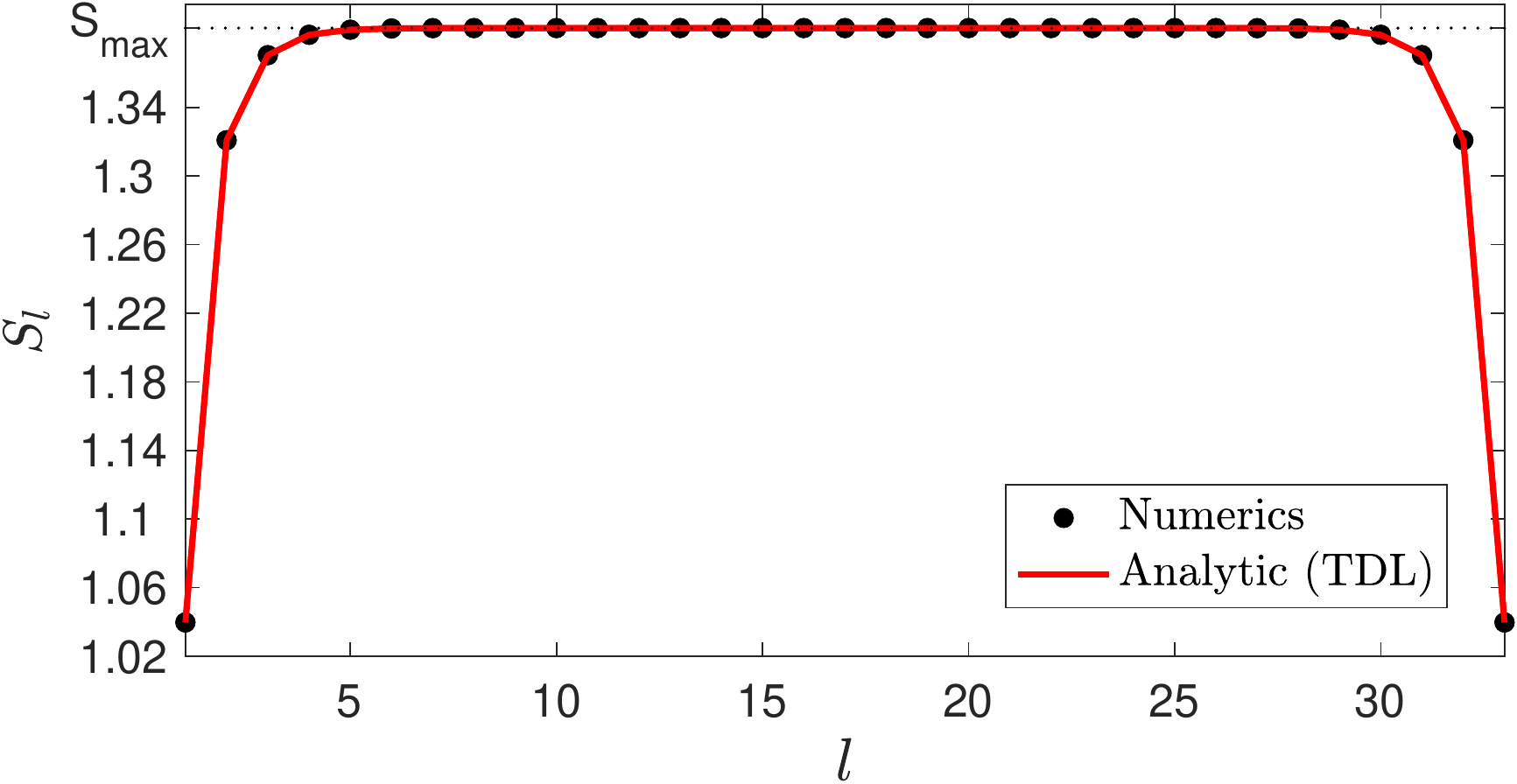}
\caption{Half-chain bipartite von Neumann Entanglement Entropy of $\ket{\psi_x}$ for $L=34$. Here, $l$ refers to the size of the subregion. Numerics shown for $l=1$ to $l=33$. The entanglement entropy is seen to saturate to $S_{max} = \log{(4)}$ (dashed line). Note this is expected from the calculations of the previous section.
As the eigenvalues of the one-site RDM (in the TDL) are $(1/4,1/4,1/2)$ (see previous section), the entanglement entropy of one site is $S = \frac{1}{2} \log{(8)} + O(e^{-L/\xi})$. Red shows the analytic expression of $S(l)$ derived in the TDL.}
  \label{fig:EECuts}
\end{figure}

The entanglement entropy of the subsystem $\mathcal{A}$ therefore admits a closed form analytic expression:
\begin{align}
S(\rho_\mathcal{A}) = -\sum_{i=1}^4 \lambda_i \log \lambda_ i
\end{align}
with $(\lambda_1, \lambda_2, \lambda_3, \lambda_4) = (1/4, 1/4, 1/4+1/2^{l+1}, 1/4-1/2^{l+1})$, which saturates to $\lim_{l \to \infty} S(\rho_\mathcal{A}) = \log(4)$.

\subsection{Numerics: Entanglement Entropy of $\ket{\psi_x}$}
\label{App:C}

While in the previous section we gave an analytic expression in the limit $L \to \infty$ for the eigenvalues of the reduced density matrix of any contiguous region (thereby also yielding the entanglement entropy), in this section we leverage the MPS representation of the state to compute, numerically, for {\it finite} $L$, its von Neumann entanglement entropy for bipartitions of the chain into two contiguous regions.
(Of course, it is expected that the differences in numerical calculations and analytic expressions will be exponentially small in system size).

In order to obtain the Schmidt decomposition of the state and thereby extract the entanglement spectrum of state, the periodic  MPS representation of $\ket{\psi_x}$ was converted into one of two \textit{open} boundary condition representations of $\ket{\psi_x}$, based on the parity of length of the subregions of the system upon bipartition, by ``doubling" the PBC MPS representation, see Fig.~\ref{fig:MPSDoubled}. 
Upon obtaining the OBC MPS representation of the state, to obtain the Schmidt decomposition of the state, we simply put the OBC representation into mixed canonical form from which we read off the singular values \cite{MPSRev}.  
Fig.~\ref{fig:EECuts} shows the half-chain von-Neumann entanglement entropy.\\

\end{appendix} 
\newpage

\bibliography{refs2}

\begin{thebibliography}{42}
\bibitem [{foo()}]{footnote}%
  \BibitemOpen
  \href@noop {} {}\bibinfo {note} {Majorizing refers to the following
  situation: suppose $|\phi\rangle$ is a pure state in a bipartite system
  $\mathcal{H}_A \otimes \mathcal{H}_B$ which has a Schmidt decomposition
  $|\phi\rangle = \sum_{n=1}^l \sqrt{\omega_n} |n\rangle_A |n\rangle_B$, where
  $l = \min(\text{dim}(\mathcal{H}_A),\text{dim}(\mathcal{H}_B))$ and
  $\sqrt{\omega_n}$ are the singular values sorted in descending order. Suppose
  $|\psi\rangle$ is another state that has corresponding Schdmit decomposition
  with singular values $\sqrt{\omega'_n}$, also sorted in descending order.
  Then for any $k=1,\cdots,l$, we say $\omega'_n$ majorizes $\omega_n$ if
  $\sum_{n=1}^k \omega_n \leq \sum_{n=1}^k \omega'_n$. In particular it means
  $S(\phi) > S(\psi)$ for any entanglement measure $S$.}\BibitemShut {Stop}%
\end{thebibliography}

\end{document}